\documentclass[prd,aps,10pt, onecolumn]{revtex4-2} 
\usepackage{natbib}
\usepackage{hyperref}
\hypersetup{colorlinks=true, citecolor=blue, urlcolor=blue, linkcolor=violet}
\usepackage{graphicx}
\usepackage{physics}
\usepackage{array}
\usepackage{amsmath,amssymb}
\usepackage{mathrsfs}
\usepackage{dsfont}
\usepackage{mathtools}
\usepackage{mathtools}
\usepackage[T1]{fontenc}
\usepackage{inputenc}
\usepackage{babel}
\usepackage{multirow}
\usepackage{longtable, array, booktabs}
\usepackage[table]{xcolor}
\usepackage{xcolor}
\usepackage{float}
\usepackage{comment}

\allowdisplaybreaks

\begin{document}
\title{Hamiltonian formulation of the $1+1$-dimensional $\phi^4$ 
theory in a momentum-space Daubechies wavelet basis
}

\author{Mrinmoy Basak}
\email{mrinmoy.263009@gmail.com}
\affiliation{Department of Theoretical Physics, Tata Institute of Fundamental Research, Mumbai 400005, India}

\author{Debsubhra Chakraborty}
\email{debsubhra@tifr.res.in}
\affiliation{Department of Theoretical Physics, Tata Institute of Fundamental Research, Mumbai 400005, India}

\author{Nilmani Mathur}
\email{nilmani@theory.tifr.res.in}
\affiliation{Department of Theoretical Physics, Tata Institute of Fundamental Research, Mumbai 400005, India}

\author{Raghunath Ratabole}
\email{ratabole@goa.bits-pilani.ac.in}
\affiliation{Department of Physics, BITS-Pilani KK Birla Goa Campus, Goa 403726, India}

\begin{abstract}
We apply the wavelet formalism of quantum field theory to investigate nonperturbative dynamics within the Hamiltonian framework. In particular, we employ Daubechies wavelets in momentum space, whose basis functions are labeled by resolution and translation indices, providing a natural nonperturbative truncation of both infrared and ultraviolet truncation of quantum field theories. As an application, we compute the energy spectra of a free scalar field theory and the interacting $1+1$-dimensional $\phi^4$ theory. This approach successfully reproduces the well-known strong-coupling phase transition in the $m^2 > 0$  regime. We find that the extracted critical coupling systematically converges toward its established value as the momentum resolution is increased, demonstrating the effectiveness of the wavelet-based Hamiltonian formulation for nonperturbative field-theoretic calculations.

\end{abstract}

\maketitle
\section{Introduction}
Although the Hamiltonian formalism remained conceptually fundamental, it was largely de-emphasized in relativistic quantum field theory during the 1950s and 1960s. This shift was driven primarily by the success of covariant Lagrangian and perturbative methods \cite{PhysRev.76.769}, as well as becasue of the absence of practical nonperturbative tools and large computational resources required for Hamiltonian diagonalization. However, between 1960 and 1980 physicists emphasized the advantages of the Hamiltonian framework \cite{PhysRev.140.B445,PhysRevD.2.1438, PhysRevD.10.2445, PhysRevD.11.395, PhysRevD.19.3715, PhysRev.125.397, PhysRev.128.2425, COLEMAN1975267, COLEMAN1976239, PhysRevD.11.2088}, because it can reveal the entire spectra, non-perturbative structure, and renormalization physics that covariant perturbation theory obscures. The revival of the Hamiltonian approach gained momentum with Wilson’s discovery of the renormalization group (RG) \cite{PhysRevB.4.3174} and the modern numerical techniques with better exact diagonalization algorithms, DMRG, and tensor-network methods. In the present age of quantum simulations, the Hamiltonian methods are seen as the best hope for {\it ab initio} computation of non-perturbative real-time physics, where Euclidean lattice fails \cite{PhysRevD.110.096016, PRXQuantum.3.020324, doi:10.1098/rsta.2021.0069, doi:10.1098/rsta.2021.0062, PhysRevA.105.032418}.

Wilson, in 1965, introduced "wave-packet" basis functions as an alternative to the prevalent Fourier or momentum basis for resolving quantum fields \cite{PhysRev.140.B445}. He proposed that these basis elements are simultaneously localized in both momentum and position space. In that work, he wanted to solve the Hamiltonian eigenvalue problem of relativistic QFTs using quantum mechanical Hamiltonian approach that go beyond perturbation theory. In order to do that, the phase space creation and annihilation operators were decomposed into localized oscillator creation and annihilation operators using a complete set of “wave packet” basis functions. He wanted to construct a set of functions, the wave-packets ($\{f(x)\})$, in such a way that the functions should rapidly goes to zero as one goes away from a certain region, i.e, in other words, the maximum energy ($\int f(x)dx$) of the functions should be accumulated within a certain region in both the position and the momentum space. However, such a wave-packet was not realized at that time. On the contrary, he constructed another set of functions, which were either fully localized in position space and had a long tail in momentum space and vice-versa, and with that he assumed the existence of these wave-packet functions.

Nearly two decades later, in 1982, Jean Morlet, a French geophysicist, along with G. Arens, E. Fourgeau, and D. Glard introduced the concept of the wavelet—a function having properties strikingly similar to the "wave-packet" functions \cite{doi:10.1190/1.1441328,doi:10.1190/1.1441329}. Subsequently,  two years later (1984), J. Morlet and Alex Grossman laid the foundation of the wavelet theory by providing the mathematical definition of the wavelet functions \cite{doi:10.1137/0515056}. One decade later, the full development of the theory was achieved through the work of Yves Meyer, Stéphane Mallat, and Ingrid Daubechies, who introduced orthonormal bases \cite{Meyer1986-1987}, multiresolution analysis \cite{Mallat1989MultiresolutionAA}, and compactly supported wavelets \cite{daubechies1992ten}, thereby establishing wavelet analysis as a mature mathematical framework with broad applications.

With the development of the compactly supported wavelets by Daubechies \cite{daubechies1992ten}, Wilson re-emphasized the importance of wavelets for non-perturbative analysis of quantum chromodynamics (QCD) in 1994 \cite{PhysRevD.49.6720}. Two types of wavelet transformations can be seen in the literature: the continuous wavelet transform (CWT) and the discrete wavelet transform (DWT). Both the approaches have been used to investigate the regularization, renormalization and gauge invariance in QFTs \cite{kessler2003waveletnotes, 10.1007/s00601-018-1357-z, michlin2017using, federbush1995new, BEST2000848, PhysRevLett.116.140403, 10.1007/JHEP06(2021)077, PhysRevD.106.036025, PhysRevD.108.125008, best1994variationaldescriptionstatisticalfield, HALLIDAY1995414, 10.1063/1.1543582, 10.1088/1751-8121/ad5503, PhysRevD.87.116011, 10.1051/epjconf/201817511002, 10.3842/SIGMA.2007.105, Albeverio2009ARO, 10.1134/S1063778818060029, PhysRevD.88.025015, 10.1007/s11182-013-9940-8, 10.1007/s10773-015-2913-7, polyzou2020lightfrontquantummechanicsquantum, PhysRevD.107.036015, PhysRevD.111.096024}. In DWT, using Daubechies wavelets, the quantum fields are resolved into discrete modes characterized by the location and the length scale. Set within the framework of multiresolution analysis \cite{10.1016/j.finel.2013.11.001}, the wavelet-based methods provide a systematic framework to improve the truncation in volume and resolution. In their 1994 work, Wilson et. al. reformulates nonperturbative QCD bound states as a weak-coupling light-front Hamiltonian problem by introducing massive constituents and applying similarity renormalization group (SRG) to achieve scale-separated, band-diagonal effective dynamics. He also advocated the use of momentum-based wavelet in the scale-separation process, first, by arranging the Hamiltonian in different scales using wavelet-based light front creation and annihilation operator, and then by applying SRG to obtain the scale-separated effective band-diagonalize Hamiltonian \cite{PhysRevD.49.6720}. However, in the subsequent period, there has been no work demonstrating the application of momentum based wavelets to the field theories. So far, in literature, the authors have focused on position space wavelet analysis of QFT. 
The original proposal of Wilson to analyse the QFTs in momentum space using wavelets, however, remains unexplored.

With this objective, as a starting point, we calculate the energy spectrum of the free scalar field theory and investigate the $1+1$ dimensional $\phi^4$ theory using Daubechies wavelets basis in the momentum space. We demonstrate the well-known strong coupling phase transition in the $m^2>0$ sector. Starting with the Hamiltonian operator in Fourier space, the momentum based creation and annihilation operators are expanded in terms of the wavelet based modes, characterized by location ($m$) and resolution   
($k$) indices. The compact support of the Daubechies wavelet basis function in Fourier space ensures that the maximum contribution to the physics comes from a significantly small number of degrees of freedom. The strength of the self-interaction and the inter-mode interaction depends on the resolution($k$), the translation index ($m$) and the order ($K$) of the Daubechies wavelets. We then construct the Fock space basis elements within this momentum-space wavelet framework, which, to the best of our knowledge, has not been reported previously in the literature. The closest related work is that of Bulut and Polyzou \cite{PhysRevD.87.116011}, who formulate relativistic quantum field theories using Daubechies wavelets in position space. In their approach, the compact support of the wavelet basis leads to inter-mode interactions that are restricted to only a finite number of neighbouring modes. However, till date no explicit field-theoretic calculations using this approach have been reported. A related study using a Fourier basis was carried out by Raychkov et al. \cite{PhysRevD.91.085011, PhysRevD.93.065014}. In standard Fourier basis approach, the interacting Hamiltonian is truncated to a finite-dimensional space of states spanned by the eigenvectors of the free Hamiltonian $H_0$, with eigenvalues below some energy cutoff $E_{max}$ \cite{PhysRevD.91.085011, PhysRevD.93.065014, 10.21468/SciPostPhys.13.2.011, 10.1007/JHEP10(2016)050, 10.1007/JHEP05(2021)190}. However, the wavelet based Fock space elements are not eiegnstates of $H_0$. Consequently, to construct the finite interacting Hamiltonian matrix, the Hamiltonian is truncated to a finite dimensional space of states spanned by the wavelet-based Fock space elements. The number of wavelet-modes of these Fock space elements are restricted in such a way that the the energy expectation values of these states do not cross $E_{max}$. Although still at a developmental stage, the advances reported in this work give us hope that such techniques will play a significant role for studying nonperturbative quantum field theory studies in future.

The flow of the paper is as follows: In Sec. \ref{sec:Daubechies_wavelet_basis}, we briefly discussed the construction of Daubecheis wavelet basis and it's associated properties, which have been used in the rest of the paper. In Sec. \ref{sec:formalism}, the formation of the Fock space basis elements and the computaion of the energy eigenvalues for the free scalar field theory within the framework of Daubechies wavelet basis is provided. In Sec. \ref{sec:the-phi_4_hamiltonian_in_Daubechies_wavelet_basis}, we detail the construction of the $\phi^4$-theory Hamiltonian matrix elements and compute the energy eigenvalues to describe the emergence of the symmetry breaking phase within $m^2>0$ sector for the strong coupling regime. The computation and the comparison with the previous studies of the critical point is also depicted in this section. The conclusion and the outlook of the work is presented in Sec. \ref{sec:conclusions_and_outlook}.
More details about various parts of calculations are provided in the Appendies at the end.

\section{Daubechies wavelet basis}
\label{sec:Daubechies_wavelet_basis}
In this section, we summarized some of the key aspects of the Daubechies wavelet basis, which we will utilize throughout the paper. More details can be found in these references \cite{PhysRevD.95.094501,kessler2003waveletnotes,PhysRevD.87.116011,https://doi.org/10.1002/cpa.3160410705,daubechies1992ten,PhysRevD.107.036015,388960,PhysRevD.111.096024}.
The Daubechies wavelets form an orthonormal basis for the space of square integrable functions, $\mathcal{L}^2(\mathbb{R})$. It has two class of constituent functions: the scaling and the wavelet function. We start by introducing the scaling functions, constructed from a single functions $s(x)$, called the mother scaling function, defined through the renormalization group equation,
\begin{eqnarray}
\label{eq:scaling_equation}
s(x) = D\!\left(
\underbrace{
\sum_{l=0}^{2K-1} h_l \, T^{\,l} s(x)
}_{\text{block average}}
\right)
\quad \text{rescale}
\end{eqnarray}
known as the \textit{scaling equation}. Here, $h_n$'s are real coefficients that can be determined from the properties of the Daubechies wavelet basis, as discussed in Ref. \cite{basak2025multiresolutionanalysisquantumtheories}. Here, $\hat{T}$ and $\hat{D}$ are translation and the contraction operation respectively. Acted upon the mother scaling function, $s(x)$, the operator $\hat{T}$ will shift the function by one unit towards the left, and $\hat{D}$ will shrink the support of the function by retaining the norm of the scaling function, $\int s(x) dx$, fixed to be $1$,
\begin{eqnarray}
D s(x)=\sqrt{2}s(2x), \quad T s(x)= s(x-1).
\end{eqnarray}
The operators $\hat{D}$ and $\hat{T}$ do not commute with each other,
\begin{eqnarray}
\label{eq:commutation_relation_D_and_T}
[\hat{D},\hat{T}]=\hat{D}\hat{T}(1-\hat{T}).
\end{eqnarray}
In the Eq. (\ref{eq:scaling_equation}), $K$, is called the order of the scaling function, which sets the maximum support of the function, i.e. $\left[0, 2K-1\right]$, and also determines the amount of smoothness of the scaling function. Throughout this paper, we have used order 3 ($K=3$) Daubechies functions. The values of $h_n$ for $K=3$ are listed in the Table. \ref{tab:h_coefficient_of_Daubechies_wavelet_basis_for_different_K}.
\begin{table}[htbp]
\begin{center}
\setlength{\tabcolsep}{1.0pc}
\newlength{\digitwidth} \settowidth{\digitwidth}{\rm 0}
\catcode`?=\active \def?{\kern\digitwidth}
\caption{$h$ coefficients of Daubechies wavelet for $K=3$.}
\label{tab:h_coefficient_of_Daubechies_wavelet_basis_for_different_K}
\vspace{1mm}
\begin{tabular}{c | c }
\specialrule{.15em}{.0em}{.15em}
\hline
$h_0$ & $\dfrac{1+\sqrt{10}+\sqrt{5}+2\sqrt{10}}{16\sqrt{2}}$ \\
$h_1$ & $\dfrac{5+\sqrt{10}+3\sqrt{5}+2\sqrt{10}}{16\sqrt{2}}$ \\ 
$h_2$ & $\dfrac{10-2\sqrt{10}+2\sqrt{5}+2\sqrt{10}}{16\sqrt{2}}$ \\ 
$h_3$ & $\dfrac{10-2\sqrt{10}-2\sqrt{5}+2\sqrt{10}}{16\sqrt{2}}$ \\
$h_4$ & $\dfrac{5+\sqrt{10}-3\sqrt{5}+2\sqrt{10}}{16\sqrt{2}}$ \\ 
$h_5$ & $\dfrac{1+\sqrt{10}-\sqrt{5}+2\sqrt{10}}{16\sqrt{2}}$ \\ 
\hline
\specialrule{.15em}{.15em}{.0em}
\end{tabular}
\end{center}
\end{table}

\begin{figure*}[htbp]
\centering
\includegraphics[scale=.61]{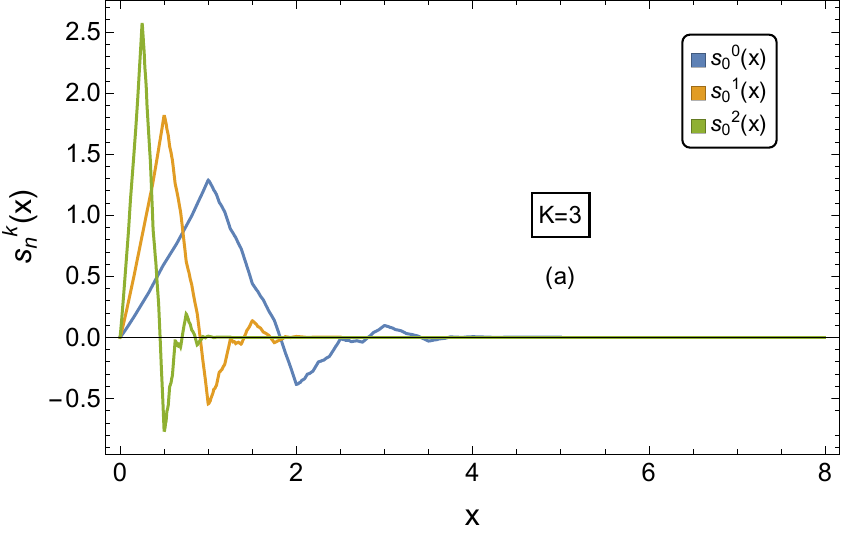}
\includegraphics[scale=0.6]{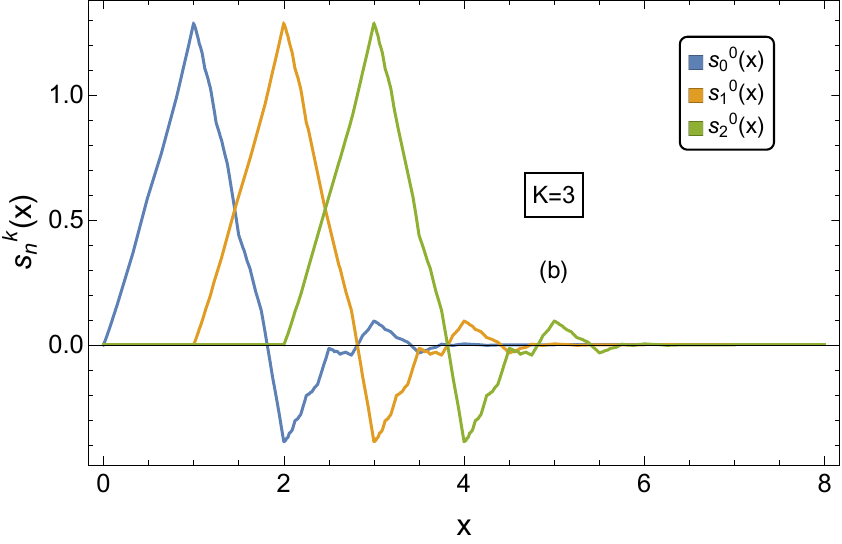}
\caption{The scaling functions for (a) different values of resolution and (b) different values of translation index are presented.}
\label{fig:the_scaling_functions_for_different_resolution_and_positions}
\end{figure*}

The solution of Eq. (\ref{eq:scaling_equation}), $s(x)$, is refered to as resolution $0$ scaling function. We can construct the resolution $k$ scaling function located at position $m$ by applying the translation operator $\hat{T}$, $m$ times and contraction operator $\hat{D}$, $k$ consecutively in the resolution $0$ scaling function, $s(x)$,
\begin{eqnarray}
\label{eq:res_k_scaling}
s^k_m(x):=\hat{D}^k\hat{T}^m s(x)=2^{\frac{k}{2}}s(2^k x-m).
\end{eqnarray}
The scaling functions for different resolutions and positions are presented in the Fig. \ref{fig:the_scaling_functions_for_different_resolution_and_positions}.

We demand the orthonormality of the scaling functions,
\begin{eqnarray}
\int s^k_m(x)s^k_n(x)dx=1.
\end{eqnarray}
The arbitrary linear combination will generate the subspace $\mathcal{H}^k$ of resolution $k$,
\begin{gather}
\mathcal{H}^k=\left\{f(x)|f(x)=\sum_{-\infty}^{\infty}f^{s,k}_n s^k_n(x),\sum_{n} |f^{s,k}_n|^2< \infty \right\}.
\end{gather}

Substituting $s(x)$ from Eq. (\ref{eq:res_k_scaling}), with the expansion of, $s(x)$ from Eq. (\ref{eq:scaling_equation}), and then using the non-commutative nature of $\hat{D}$ and $\hat{T}$ as given in Eq. (\ref{eq:commutation_relation_D_and_T}), we arrive at the following relation,
\begin{eqnarray}
s^{k}_n(x)=\sum_{m=2n}^{2n+2K-1}H_{nm}s^{k+1}_m(x),\quad H_{nm}=h_{m-2n}.
\end{eqnarray}
This equation depicts the construction of resolution $k$ scaling function from resolution $k+1$ scaling function. From this relation, it can also be inferred that the resolution $k$ subspace is contained within resolution $k+1$ subspace,
\begin{eqnarray}
\mathcal{H}^{k}\subset \mathcal{H}^{k+1}.
\end{eqnarray}
By induction, for $m>0$
\begin{eqnarray}
\mathcal{H}^{k}\subset \mathcal{H}^{k+m}.
\end{eqnarray}
These subspaces have a nested structure and the infinite resolution limit will generate the space of square integrable functions,
\begin{eqnarray}
\mathcal{L}^2(\mathbb{R})=\lim_{k\to \infty}\mathcal{H}^k.
\end{eqnarray}
Any square integrable functions hence can be expanded in this basis using only the scaling functions,
\begin{eqnarray}
\label{eq:expansion_of_square_integrable_function_in_terms_of_scaling_function}
f(x)=\lim_{k\to \infty}\sum_{n=-\infty}^{\infty}f^{s,k}_n s^k_n(x),
\end{eqnarray}
such that,
\begin{eqnarray}
\lim_{k\to \infty}\sum_{n=-\infty}^{\infty}|f^{s,k}_n|^2< \infty.
\end{eqnarray}

\begin{figure*}[htbp]
\centering
\includegraphics[scale=0.6]{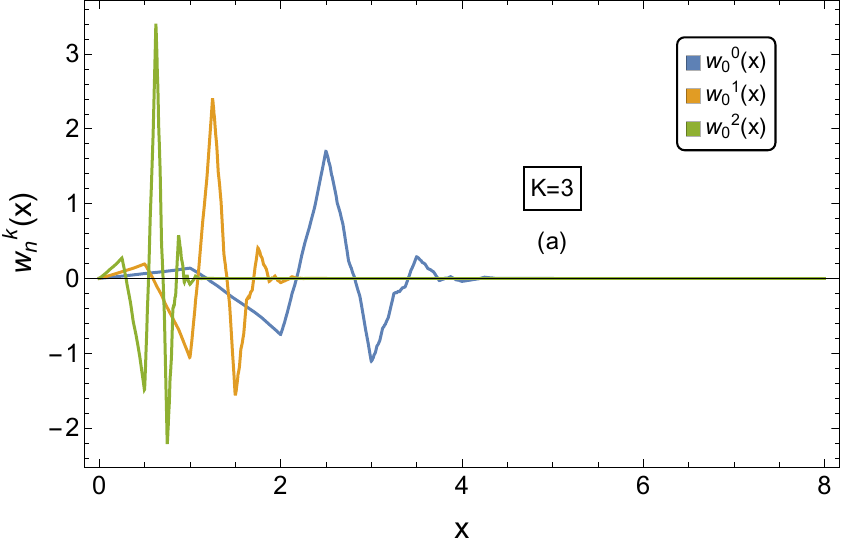}
\includegraphics[scale=0.61]{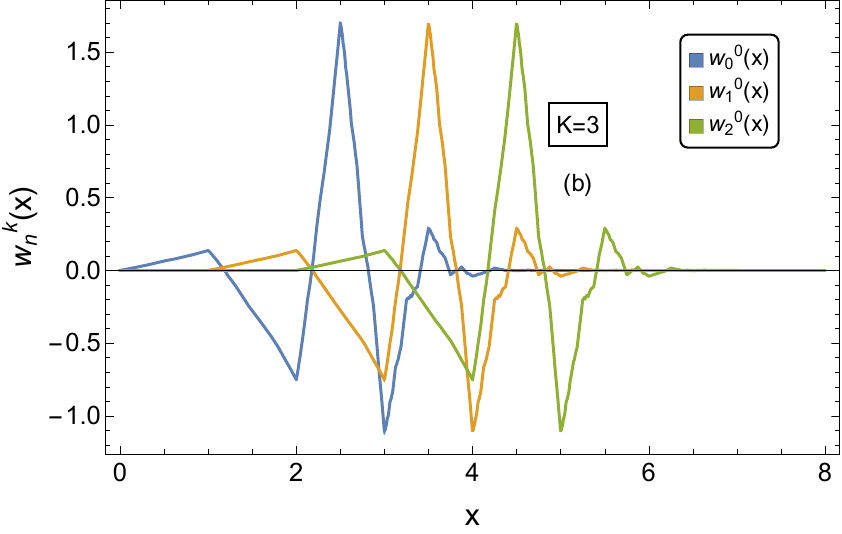}
\caption{The wavelet functions for (a) different values of resolution and (b) different values of translation index are presented.}
\label{fig:the_wavelet_functions_for_different_resolution_and_positions}
\end{figure*}

Since the resolution $k$  subspace, denoted as $\mathcal{H}^{k}$, is a proper subspace of the resolution $k+1$ subspace, $\mathcal{H}^{k+1}$, we can define another subspace, $\mathcal{W}^k$, which is orthogonal to $\mathcal{H}^{k}$ within $\mathcal{H}^{k+1}$. This means that $\mathcal{H}^{k+1}$ can be expressed as the direct sum of the resolution $k$ subspace and its orthogonal complement referred to as the wavelet subspace,
\begin{eqnarray}
\label{eq:formation_of_res_k+1_space}
\mathcal{H}^{k+1}=\mathcal{H}^k\oplus \mathcal{W}^k.
\end{eqnarray}
The basis elements of the subspace, $\mathcal{W}^k$, are the second constituent of the Daubechies wavelet basis, known as the wavelet function. These basis elements can be obtained from a single function, $w$, referred to as mother wavelet function, defined through the following equation, 
\begin{eqnarray}
\label{eq:mother_wavelet_equation}
w(x):=\sum_{k=0}^{2K-1}g_n \hat{D}\hat{T}s(x),
\end{eqnarray}
where,
\begin{eqnarray}
g_n=(-1)^n h_{2K-1-n}.
\end{eqnarray}
The coefficients, $g_n$, are chosen in such a way that the functions, $w(k)$, will be orthogonal to $s(x)$. Similar to the case of the scaling function, the resolution $k$ wavelet function, situated at the position $m$ can be constructed by applying the contraction operator, $\hat{T}$, $m$ times followed by the translation operator $\hat{D}$, $k$ times,
\begin{eqnarray}
w^k_m(x):= \hat{D}^k\hat{T}^m w(x).
\end{eqnarray}
The wavelet functions for different resolutions and positions are presented in the Fig. \ref{fig:the_wavelet_functions_for_different_resolution_and_positions}.

By design, they are orthogonal to the scaling functions,
\begin{eqnarray}
\int s^k_m(x)w^l_n(x)dx = 0,
\end{eqnarray}
and they construct an orthonormal set,
\begin{eqnarray}
\int w^k_m(x)w^l_n(x)dx=\delta_{kl}\delta_{mn}.
\end{eqnarray}
The resolution $k$ wavelet functions spans the resolution $k$ wavelet space $\mathcal{W}^k$,
\begin{gather}
\mathcal{W}^k=\left\{f(x)|f(x)=\sum_{-\infty}^{\infty}f^{w,k}_n w^k_n(x),\sum_{n} |f^{w,k}_n|^2< \infty \right\}.
\end{gather}
Recursive application of the Eq. (\ref{eq:formation_of_res_k+1_space}) will create the space of the square integrable functions, $\mathcal{L}^2(\mathbb{R})$,
\begin{eqnarray}
\mathcal{L}^2(\mathcal{R})= \mathcal{H}^{k}\oplus \mathcal{W}^{k} \oplus \mathcal{W}^{k+1} \oplus \mathcal{W}^{k+2}...,
\end{eqnarray}

Note a square integrable function can also be expanded in terms of the scaling and wavelet functions,
\begin{eqnarray}
f(x)=\sum_{n=-\infty}^{\infty}f^{s,k}_n s^k_n(x)+\sum_{n=-\infty}^{\infty}\sum_{l=k}^{\infty} f^{w,l}_n w^l_n(x).
\end{eqnarray}
such that,
\begin{eqnarray}
\sum_{n=-\infty}^{\infty}|f^{s,k}_n|^2 + \sum_{n=-\infty}^{\infty}\sum_{l=k}^{\infty} |f^{w,l}_n|^2<\infty.
\end{eqnarray}

From the above discussion, we can infer that there are two possible choice of the basis functions for the truncated subspace of resolution $k$, $\mathcal{H}^k$: one can either work only with the scaling functions $\left\{s^k_m(x)\right\}_{n=-\infty}^{\infty}$ of resolution $k$ or else one can work with the combination of both the scaling and the wavelet functions of resolution $k-1$, $\left(\left\{s^{k-1}(x)\right\}_{n=-\infty}^{\infty}\cup\left\{w^{k-1}(x)\right\}_{n=-\infty}^{\infty}\right) $ and there are orthogonal transformations which can relate the two basis functions:
\begin{gather}
s^{k-1}_n(x)=\sum_{m=2n}^{2n+2K-1}H_{nm}s^k_m(x),\\
w^{k-1}_n(x)=\sum_{m=2n}^{2n+2K-1}G_{nm}w^k_m(x),\\
s^k_n(x)=\sum_{m=2n}^{2n+2K-1}H^t_{nm}s^{k-1}_m(x)+\sum_{m=2n}^{2n+2K-1}G^t_{nm}w^{k-1}_n(x),
\end{gather}
where
\begin{gather}
H_{nm}=h_{m-2n},\quad G_{nm}=g_{m-2n},\\
H^t_{nm}=h_{n-2m},\quad G^t_{nm}=g_{n-2m}.
\end{gather}

Since Daubechies wavelets have compact support, the Fourier transformmation of the basis functions does not have compact support, they  are infinitely differentiable and fall off like inverse powers of the Fourier variable.

\section{Wavelet-based Fock space}
\label{sec:formalism}
In this section, using the example of the $1+1$ dimensional free scalar field theory, we present a construction of the wavelet-based Fock space. We then demonstrate the explicit construction of the Hamiltonian matrix elements in this Fock basis, followed by the truncation scheme employed to obtain a finite-dimensional Hamiltonian matrix. The resulting finite Hamiltonian is subsequently diagonalized to compute the energy eigenvalues at different resolutions. The analysis in this section shows that wavelets provide a consistent methodology for the analysis of Hamiltonian QFTs.
\subsection{Scalar Field Theory: Fourier Modes}
The Lagrangian density for the $1+1$ dimensional scalar field theory with unit bare mass is given by,
\begin{eqnarray}
\mathcal{L}_{0}=\frac{1}{2}\left[\left(\frac{\partial \phi}{\partial t}\right)^2-\left(\frac{\partial \phi}{\partial t}\right)^2+ \phi^2\right].
\end{eqnarray}
The corresponding normal ordered Hamiltonian is,
\begin{eqnarray}
\label{eq:scalar_field_theory_Hamiltonian}
\textrm{H}_0=\int dx \frac{1}{2}{\rm N}\left[\pi^2+\left(\frac{\partial \phi}{\partial x}\right)^2 + \phi^2\right],
\end{eqnarray}
where,
\begin{eqnarray}
\pi=\frac{\partial \mathcal{L}}{\partial \dot{\phi}}=\dot{\phi}.
\end{eqnarray}
While working in the Schrödinger representation, we expand the field operator, $\phi (x)$, and its canonical conjugate, $\pi (x), $ in the Fourier basis,
\begin{gather}
\label{eq:expansion_of_phi_x_in_momentum_modes}
\phi(x)=\int \frac{dp}{\sqrt{2(2\pi)2E}}\left(e^{-ipx}a(p)+e^{ipx}a^{\dagger}(p)\right),\\
\label{eq:expansion_of_pi_x_in_momentum_modes}
\pi(x)=\int \frac{(iE)dp}{\sqrt{2(2\pi)2E}}\left(e^{ipx}a^{\dagger}(p)-e^{-ipx}a(p)\right).
\end{gather}
Substituting Eq. (\ref{eq:expansion_of_phi_x_in_momentum_modes}) and (\ref{eq:expansion_of_pi_x_in_momentum_modes}) in Eq. (\ref{eq:scalar_field_theory_Hamiltonian}), we get the following momentum space representation of the normal-ordered Hamiltonian,
\begin{eqnarray}
\label{eq:scalar_field_Hamiltonian_operator_momentum_space_expansion}
\textrm{H}_0 =\int dp E_p a^{\dagger}(p)a(p).
\end{eqnarray}
\subsection{Scalar Field Theory: Wavelet Modes}
\label{sec:the_momentum_spce_wavelet_representation_of_scalar_field_theory}
We now present the scalar field theory Hamiltonian, Eq. (\ref{eq:scalar_field_Hamiltonian_operator_momentum_space_expansion}), in terms of the scaling mode creation and annihilation operator. This is done by expanding the momentum space creation operator, $a^{\dagger}(p)$ and annihilation operator, $a(p)$, in terms of scaling bases. The basis coefficients represent the scaling-mode creation and annihilation operators labelled by resolution and location indices, as described in Eq. (\ref{eq:expansion_of_square_integrable_function_in_terms_of_scaling_function}),
\begin{eqnarray}
\label{eq:creation_operator_scaling_mode_expansion}
a^{\dagger}(p)&=&\lim_{k\to \infty}\sum_{n=-\infty}^{\infty}a^{s,k\dagger}_n s^k_n(p),\\
\label{eq:annihilation_operator_scaling_mode_expansion}
a(p)&=&\lim_{k\to \infty}\sum_{n=-\infty}^{\infty}a^{s,k}_n s^k_n(p).
\end{eqnarray}
The scaling-mode creation and annihilation operators are given by the inverse of Eq. (\ref{eq:creation_operator_scaling_mode_expansion}), and (\ref{eq:annihilation_operator_scaling_mode_expansion}), respectively,
\begin{eqnarray}
    a^{s,k\dagger}_n &=& \int a^{\dagger}(p) s^k_n(p) dp,\\
     a^{s,k}_n &=& \int a(p) s^k_n(p) dp.
\end{eqnarray}
When acted upon the vacuum, the creation operator does not create a definite momentum state but creates a state, smeared over a momentum range $\frac{2K-1}{2^k}$,  starting form $n$.
Due to the compact support of the scaling function, each momentum mode can be constructed using a finite number of scaling modes. For example, while working with the order, $K=3$, Daubechies wavelet basis, the momentum $0$ mode ($p=0$) can be approximately constructed using the linear combination of four resolution $0$ scaling modes with translation index $-4\leq n\leq 0$.

\begin{figure}[hbpt]
\centering
\includegraphics[scale=0.5]{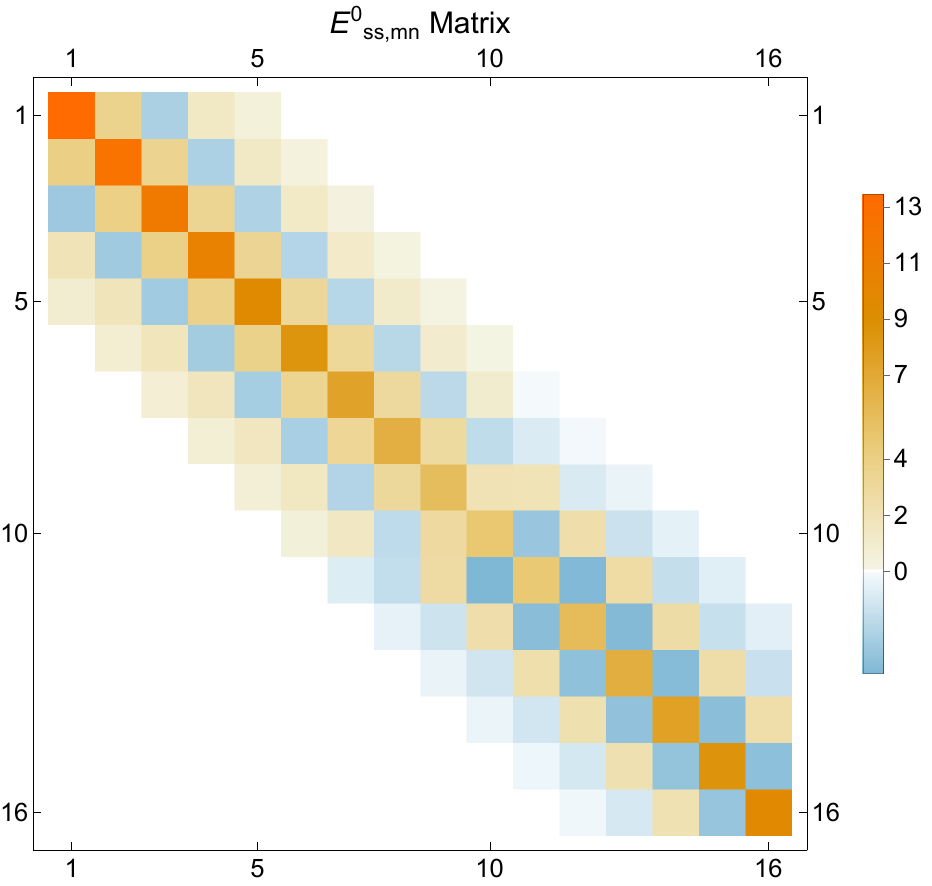}
\caption{The hopping strength between two different scaling-modes located at two positions are presented.}
\label{fig:the_hopping_strength}
\end{figure}

Substituting Eq. (\ref{eq:creation_operator_scaling_mode_expansion}) and (\ref{eq:annihilation_operator_scaling_mode_expansion}) into expansions Eq. (\ref{eq:scalar_field_Hamiltonian_operator_momentum_space_expansion}), we get the following form of the Hamiltonian,
\begin{eqnarray}
\label{eq:the_Hamiltonian_Daubechies_basis}
\mathrm{H}_0=\lim_{k\to \infty}\sum_{m,n}E^k_{ss,mn}a^{s,k\dagger}_ma^{s,k}_n,
\end{eqnarray}
where,
\begin{eqnarray}
\label{eq:hopping_strength}
E^k_{ss,mn}=\int \sqrt{p^2+1}s^k_m(p)s^k_n(p)dp.
\end{eqnarray}
Eq.  (\ref{eq:the_Hamiltonian_Daubechies_basis}) shows that the scaling modes can hop between positions $m$ and $n$, with the hopping strength given by the above overlap integral (Eq. \ref{eq:hopping_strength}). Due to the compact support of the scaling function, only a few of these integrals have a non-zero value for a given value of $m$. For example, $E^k_{ss,0n}$ is non-zero only for $-4\leq n\leq 4$. Fig. (\ref{fig:the_hopping_strength}) shows that the hopping strength decreases as the difference between the translation indices of two scaling modes, $|m-n|$, increases, and eventually it becomes zero. This implies that the locality of the theory will be preserved within this framework.

Any generic multi-particle Fock basis state in this framework can be defined as,
\begin{eqnarray}
\label{eq:scaling_mode_fock_basis}
\ket{m1^{s,k}_{-\infty},\ldots, m1^{s,k}_p,\ldots, m1^{s,k}_{\infty}}.
\end{eqnarray}
Here, $m1^{s,k}_p$ is the number of resolution-$k$ scaling modes located at position $p$. The action of the scaling creation operator, $a^{s,k\dagger}_n$, and the scaling annihilation operator, $a^{s,k}_n$, on this state  is given by,
\begin{widetext}
\begin{gather}
\label{eq:action_a_dagger_on_fock_basis}
a^{s,k\dagger}_n\ket{m1^{s,k}_{-\infty},\ldots, m1^{s,k}_n,\ldots, m1^{s,k}_{\infty}}=\sqrt{m1^{s,k}_n+1}\ket{m1^{s,k}_{-\infty},\ldots, m1^{s,k}_n+1,\ldots, m1^{s,k}_{\infty}},\\
\label{eq:action_a_on_fock_basis}
a^{s,k}_n\ket{m1^{s,k}_{-\infty},\ldots, m1^{s,k}_n,\ldots, m1^{s,k}_{\infty}}=\sqrt{m1^{s,k}_n}\ket{m1^{s,k}_{-\infty},\ldots, m1^{s,k}_n-1,\ldots, m1^{s,k}_{\infty}}
\end{gather}  
Using Eq. (\ref{eq:the_Hamiltonian_Daubechies_basis}), (\ref{eq:action_a_dagger_on_fock_basis}), and (\ref{eq:action_a_on_fock_basis}), the matrix elements of the Hamiltonian operator in the wavelet-based Fock basis can hence be calculated leading to,
\begin{eqnarray}
\mathrm{H}_{0ss,m1m2}&&=\bra{m1^{s,k}_{-\infty},\ldots,m1^{s,k}_{\infty}}\mathrm{H}_0\ket{m2^{s,k}_{-\infty},\ldots, m2^{s,k}_{\infty}},\nonumber\\
&&
\label{eq:free_hamiltonian_matrix_element}
\begin{cases}
=\sum_{p,q}T^{k}_{ss,pq}\delta_{m1^{s,k}_{-\infty},m2^{s,k}_{-\infty}}\ldots \delta_{m1^{s,k}_{\infty},m2^{s,k}_{\infty}},\,\, \text{for}\,\, p= q\\
=\sum_{p,q}T^{k}_{ss,pq}\delta_{m1^{s,k}_{-\infty},m2^{s,k}_{-\infty}}\ldots \delta_{m1^{s,k}_p,m2^{s,k}_p+1}\ldots \delta_{m1^{s,k}_q,m2^{s,k}_q-1}\ldots \delta_{m1^{s,k}_{\infty},m2^{s,k}_{\infty}},\,\, \text{for}\,\, p\neq q ,
\end{cases}
\end{eqnarray}
where,
\begin{eqnarray}
T^{k}_{ss,pq}=
\begin{cases}
E^{k}_{ss,pp}m2^{s,k}_p \,\, \text{for}\,\, p=q\\
E^{k}_{ss,pp}\sqrt{(m2^{s,k}_p+1)m2^{s,k}_q}\,\,\text{for}\,\, p\neq q .
\end{cases}
\end{eqnarray}
\end{widetext}

\subsection{Hamiltonian Eigenvalues}
The finite Hamiltonian matrix is obtained by constraining the maximum value of allowed average momentum of each of the wavelet modes, and the maximum allowed average energy of each of the Fock space basis elements constructed using those modes. The restriction on the maximum allowed average momentum of each particle is imposed by truncating the momentum space volume, which in turn restricts the translation index. The maximum allowed average energy of the system is constrained by restricting the total particle number of each state in such a way that the average energy of that particular state does not exceed the maximum allowed value of the energy. Constraining the resolution amounts to putting the volume cutoff on the real space, which is determined by the Fourier transformation of the basis functions. We use the periodic boundary condition to carry out these computations. In this simulation, the maximum value of both the average momentum and energy are restricted by $10$.  The eigenvalues are computed by diagonalizing the finite Hamiltonian matrix. The eigenvalues thus obtained, for the first few resolutions, are presented in the Table. \ref{tab:the_eigenvalues_of_h_increasing_resolution_only_scaling}. 

\begin{table}[hbpt]
\centering
\setlength{\tabcolsep}{.8pc}
\caption{The zero momentum eigenvalues of the truncated Hamiltonian, which is constructed with the Fock-space basis elements comprises of only scaling particles, corresponding to the one, two, three, and four particle ground states with increasing resolution.}
\label{tab:the_eigenvalues_of_h_increasing_resolution_only_scaling}
\vspace{1mm}
\begin{tabular}{c c c c c c}
\specialrule{.15em}{.0em}{.15em}
\hline
states & Exact & $k=0$ 	    &   $k=1$   &   $k=2$ 	 \\
\hline
$1p$    & $1$  & $1.04363$  & $1.01152$ & $1.00408$    \\
$2p$ 	& $2$  & $2.08726$  & $2.02304$ & $2.00815$    \\
$3p$ 	& $3$  & $3.13093$  & $3.03457$ & $3.01223$    \\
$4p$ 	& $4$  & $4.17466$  & $4.04613$ & $4.01630$    \\
\hline
\specialrule{.15em}{.15em}{.0em}
\end{tabular}
\end{table}

As can be seen from Table \ref{tab:the_eigenvalues_of_h_increasing_resolution_only_scaling}, the eigenvalues are approaching towards the exact value with the increasing resolution. This promising convergence pattern hence motivate us to apply this formalism further to the $\phi^4$-theory, which we dicuss below.

\section{The $\phi^4$ Hamiltonian in Daubechies wavelet based framework}
\label{sec:the-phi_4_hamiltonian_in_Daubechies_wavelet_basis}
Following the same approach presented in the previous section, in this section, we employ the same techniques and investigate the $1+1$ dimensional $\phi^4$ theory with an interactive term. Then, we present the expansion of the interaction term in terms of wavelet-based creation and annihilation operators. Subsequently, the construction of the interaction Hamiltonian matrix elements in the Fock-space basis is demonstrated. The computed energy eigenvalues for different values of coupling strength $\lambda$ is also listed. Finally, the emergence of the symmetry breaking phase and the critical point for the symmetry breaking is presented.

\subsection{Interactive $\phi^4$-term: Fourier Modes}
The additive interactive term of the $1+1$ dimensional $\phi^4$-theory Hamiltonian is given by,
\begin{eqnarray}
\mathrm{H_I}=\frac{\lambda}{4!}\int dx \phi^4, \quad \lambda>0.
\end{eqnarray}
Using the expansion of $\phi(x)$ in terms of the momentum modes, given in Eq. (\ref{eq:expansion_of_phi_x_in_momentum_modes}), we reach to the following expression of the normal ordered interacting Hamiltonian, in terms of the momentum creation and annihilation operator as,
\begin{widetext}
\begin{gather}
\mathrm{H_I}=\frac{\lambda}{4!}\int_{p_1+p_2+p_3+p_4=0}\frac{dp_1 dp_2 dp_3 dp_4}{2^2(2\pi)^2\sqrt{E_1E_2E_3E_4}}\left[a(-p_1)a(-p_2)a(-p_3)a(-p_4)\right.\nonumber\\
\left.
-4\times a^{\dagger}(p_1)a(-p_2)a(-p_3)a(-p_4) \right.\nonumber\\
\left. +6\times a^{\dagger}(p_1)a^{\dagger}(p_2)a(-p_3)a(-p_4)- 4\times a^{\dagger}(p_1)a^{\dagger}(p_2)a^{\dagger}(p_3)a(-p_4)+a^{\dagger}(p_1)a^{\dagger}(p_2)a^{\dagger}(p_3)a^{\dagger}(p_4)\right]
\end{gather}
\subsection{Interactive $\phi^4$-term: Wavelet Modes}
Substituting the expansions of momentum mode creation and annihilation operator in terms of the scaling mode creation and annihilation operator, as shwon in Eqs. (\ref{eq:creation_operator_scaling_mode_expansion}) and (\ref{eq:annihilation_operator_scaling_mode_expansion}), we arrive at the following expression of the interaction Hamiltonian,
\begin{gather}
\mathrm{H_I}=\frac{\lambda}{4!}\sum_{q_1,q_2,q_3,q_4}\frac{\Gamma^k_{q_1,q_2,q_3,q_4}}{(4\pi)^2}\left[a^{s,k}_{q_1}a^{s,k}_{q_2}a^{s,k}_{q_3}a^{s,k}_{q_4}-4\times a^{s,k\dagger}_{q_1}a^{s,k}_{q_2}a^{s,k}_{q_3}a^{s,k}_{q_4}\right.\nonumber\\
\left.
+6\times a^{s,k\dagger}_{q_1}a^{s,k\dagger}_{q_2}a^{s,k}_{q_3}a^{s,k}_{q_4}-4\times a^{s,k\dagger}_{q_1}a^{s,k\dagger}_{q_2}a^{s,k\dagger}_{q_3}a^{s,k}_{q_4}\right.\nonumber\\
\label{eq:interacting_hamiltonian_scaling_particl_mode}
\left.+a^{s,k\dagger}_{q_1}a^{s,k\dagger}_{q_2}a^{s,k\dagger}_{q_3}a^{s,k\dagger}_{q_4}\right],
\end{gather}
where,
\begin{gather}
\label{eq:phi4_hopping_term}
\Gamma^k_{q_1,q_2,q_3,q_4}=\int_{p_1+p_2+p_3+p_4=0}\frac{dp_1dp_2dp_3dp_4}{\sqrt{E_1E_2E_3E_4}}s^k_{q_1}(p_1)s^k_{q_2}(p_2)s^k_{q_3}(p_3)s^k_{q_4}(p_4),\quad\quad E_r=p_r^2+m^2.
\end{gather}
The procedure to efficiently calculate these overlap integrals are detailed in the Appendix \ref{appen:phi_4_overlap_integral}.
\end{widetext}
\subsection{Hamiltonian ($\mathrm{H}_0$+$\mathrm{H_I}$) Eigenvalues: the critical point}
The Hamiltonian matrix element within the Daubechies wavelet based framework is obtained by inserting the interacting Hamiltonian operator, Eq. (\ref{eq:interacting_hamiltonian_scaling_particl_mode}), in between the scaling Fock space basis states (Eq. (\ref{eq:scaling_mode_fock_basis})). As in the case of the Hamiltonian matrix elements of the free scalar field (Eq. (\ref{eq:free_hamiltonian_matrix_element})), applying the procedure outlined in Sec. III leads to the following expression for the Hamiltonian matrix elements,
\begin{widetext}
\begin{gather}
\mathrm{H_I}_{ss,m_1 m_2}= \frac{\lambda}{4!}\sum_{q_1,q_2,q_3,q_4}\frac{\Gamma^k_{q_1,q_2,q_3,q_4}}{(4\pi)^2}\left[\mathrm{A}^{k}_{ss,m_1m_2}-4 \times \mathrm{B}^{k}_{ss,m_1m_2}+6\times \mathrm{C}^{k}_{ss,m_1m_2}- 4 \times \mathrm{D}^{k}_{ss,m_1m_2} + \mathrm{E}^{k}_{ss,m_1m_2} \right],
\end{gather}
where,
\begin{gather}
\begin{cases}
 \mathrm{A}^{k}_{ss,m1m2}=\bra{m1^{s,k}_{-\infty}...m1^{s,k}_{\infty}}a^{s,k}_{q_1}a^{s,k}_{q_2}a^{s,k}_{q_3}a^{s,k}_{q_4}\ket{m2^{s,k}_{-\infty}...m2^{s,k}_{\infty}}\\
 \mathrm{B}^{k}_{ss,m_1m_2}=\bra{m1^{s,k}_{-\infty}...m1^{s,k}_{\infty}}a^{s,k\dagger}_{q_1}a^{s,k}_{q_2}a^{s,k}_{q_3}a^{s,k}_{q_4}\ket{m2^{s,k}_{-\infty}...m2^{s,k}_{\infty}}\\
 \mathrm{C}^{k}_{ss,m_1m_2}=\bra{m1^{s,k}_{-\infty}...m1^{s,k}_{\infty}}a^{s,k\dagger}_{q_1}a^{s,k\dagger}_{q_2}a^{s,k}_{q_3}a^{s,k}_{q_4}\ket{m2^{s,k}_{-\infty}...m2^{s,k}_{\infty}}\\
 \mathrm{D}^{k}_{ss,m_1m_2}=\bra{m1^{s,k}_{-\infty}...m1^{s,k}_{\infty}}a^{s,k\dagger}_{q_1}a^{s,k\dagger}_{q_2}a^{s,k\dagger}_{q_3}a^{s,k}_{q_4}\ket{m2^{s,k}_{-\infty}...m2^{s,k}_{\infty}}\\
\mathrm{E}^{k}_{ss,m_1m_2}=\bra{m1^{s,k}_{-\infty}...m1^{s,k}_{\infty}}a^{s,k\dagger}_{q_1}a^{s,k\dagger}_{q_2}a^{s,k\dagger}_{q_3}a^{s,k\dagger}_{q_4}\ket{m2^{s,k}_{-\infty}...m2^{s,k}_{\infty}} .
 \end{cases}
 \end{gather}
The evaluation of these terms is discussed in the appendix. \ref{appen:the_interacting_Hamiltonian_matrix}
\end{widetext}
\begin{figure*}[H]
\centering
\includegraphics[scale=0.58]{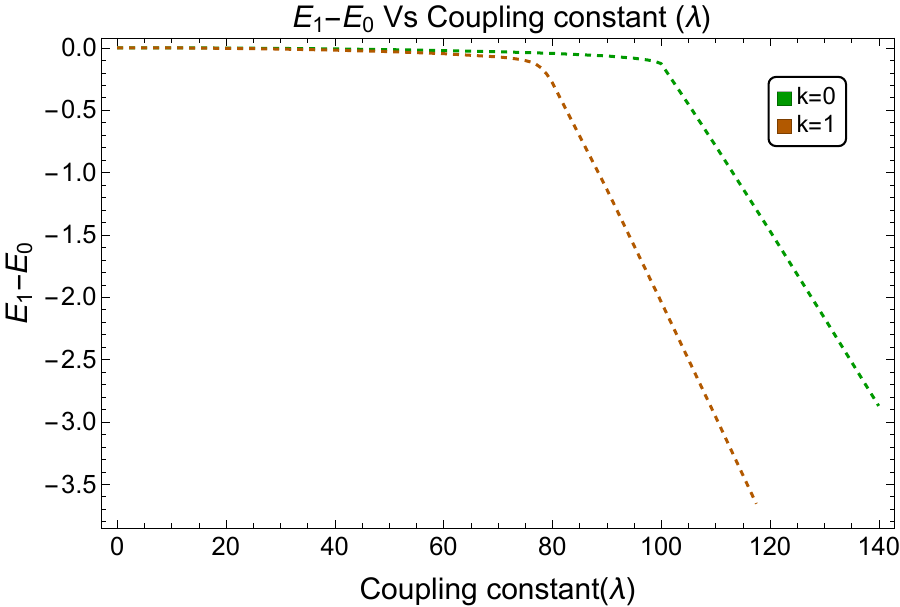}
\label{fig:The_divergence_of_ground_state}
\includegraphics[scale=0.58]{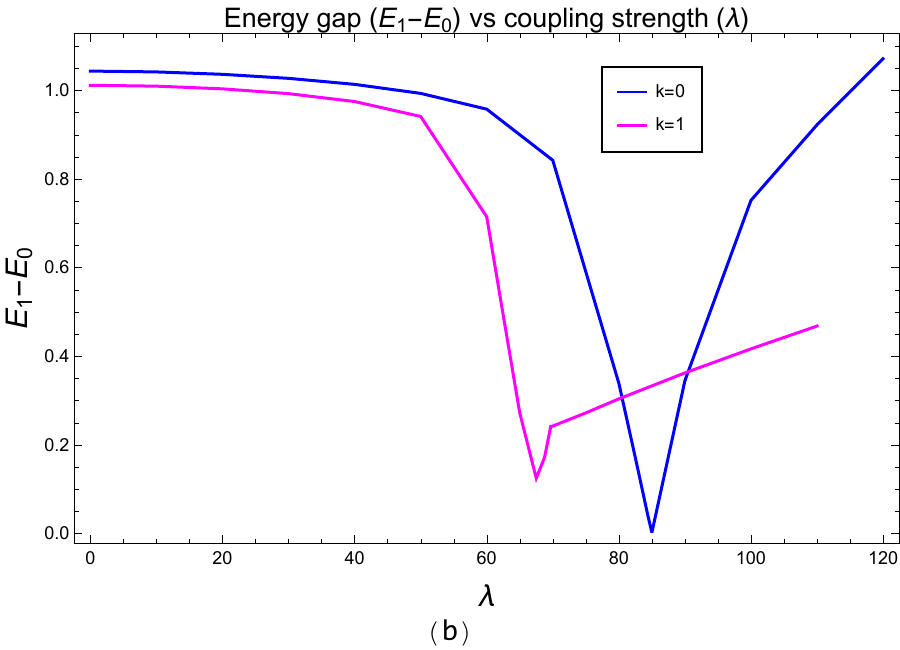}
\label{fig:The_degeneracy_of_ground_state}
\caption{(a) The divergence of the ground state energy with the increasing value of the coupling constant $\lambda$, (b) the variation of the first excited state with the increasing value of the coupling constant $\lambda$.}
\end{figure*}

For a fixed value of the coupling constant $\lambda$, the finite Hamiltonian matrix is constructed by evaluating the matrix elements of the full-Hamiltonian (free-field part plus interaction term) within a truncated Fock-space subspace which is obtained using the truncation procedure described in Sec.~\ref{sec:the_momentum_spce_wavelet_representation_of_scalar_field_theory}. For the successively increasing value of $\lambda$, the finite Hamiltonian matrix is then diagonalized to calculate the energy eigenvalues of the system, and the results are presented in Table. \ref{tab:eigenvalues_phi_4_theory_different_values_of_lambda}.
\begin{table*}[htbp]
\begin{center}
\caption{The energy eigenvalues of the $1+1$ dimensional $\phi^4$ theory for resolutions $1$ and $2$ and for the increasing value of the coupling constant $\lambda$.}
\label{tab:eigenvalues_phi_4_theory_different_values_of_lambda}
\setlength{\tabcolsep}{0.5pc}
\vspace{1mm}
\begin{tabular}{c | c | c | c | c | c | c | c | c}
\specialrule{.15em}{.0em}{.15em}
\hline
\multirow{2}{*}{$\lambda$} & \multicolumn{4}{c|}{$k=0$} & \multicolumn{4}{c}{$k=1$}\\
\cline{2-9}
 & $E_0$ & $E_1$ & $E_2$ & $E_3$ & $E_0$ & $E_1$ & $E_2$ & $E_3$\\
\hline
0 & 0.00000 & 1.04351 & 1.41908 & 1.45087 & 0.00000 & 1.01149 & 1.12243 & 1.13259 \\
10 & -0.00058 & 1.04111 & 1.41720 & 1.44918 & -0.00115 & 1.00844 & 1.11967 & 1.13001 \\
20 & -0.00231 & 1.03413 & 1.41143 & 1.44426 & -0.00464 & 0.99923 & 1.11115 & 1.12239 \\
30 & -0.00522 & 1.02256 & 1.40120 & 1.43623 & -0.01056 & 0.98323 & 1.09579 & 1.10984 \\
40 & -0.00940 & 1.00588 & 1.38506 & 1.42525 & -0.01919 & 0.95863 & 1.07106 & 1.09226 \\
50 & -0.01500 & 0.98277 & 1.35983 & 1.41150 & -0.03105 & 0.92062 & 1.03085 & 1.06864 \\
60 & -0.02231 & 0.95024 & 1.31787 & 1.39504 & -0.04740 & 0.85115 & 0.95189 & 1.03579 \\
70 & -0.03186 & 0.90016 & 1.23400 & 1.37542 & -0.07340 & 0.58345 & 0.62606 & 0.70603 \\
80 & -0.04494 & 0.80139 & 1.01208 & 1.27131 & -0.27996 & -0.12752 & -0.00315 & 0.03254 \\
90 & -0.06562 & 0.47094 & 0.63609 & 0.79445 & -1.13289 & -0.88706 & -0.78133 & -0.59739 \\
100 & -0.12726 & -0.11981 & 0.17024 & 0.24067 & -2.03649 & -1.74348 & -1.60327 & -1.38075 \\
110 & -0.78552 & -0.46776 & -0.40604 & -0.25310 & -2.95791 & -2.62094 & -2.44741 & -2.18812 \\
120 & -1.46849 & -1.08937 & -1.02572 & -0.80018 & -3.89041 & -3.51085 & -3.30540 & -3.00996 \\
130 & -2.16576 & -1.74832 & -1.66658 & -1.37816 & -4.83051 & -4.40916 & -4.17293 & -3.84155 \\
140 & -2.87248 & -2.42234 & -2.32014 & -1.96964 & -5.77612 & -5.31350 & -5.04738 & -4.68012 \\
150 & -3.58589 & -3.10588 & -2.98226 & -2.56988 & -6.72585 & -6.22233 & -5.92707 & -5.52393 \\
\hline
\specialrule{.15em}{.15em}{.0em}
\end{tabular}
\end{center}
\end{table*}

\begin{figure}[hbtp]
\centering
\includegraphics[scale=0.55]{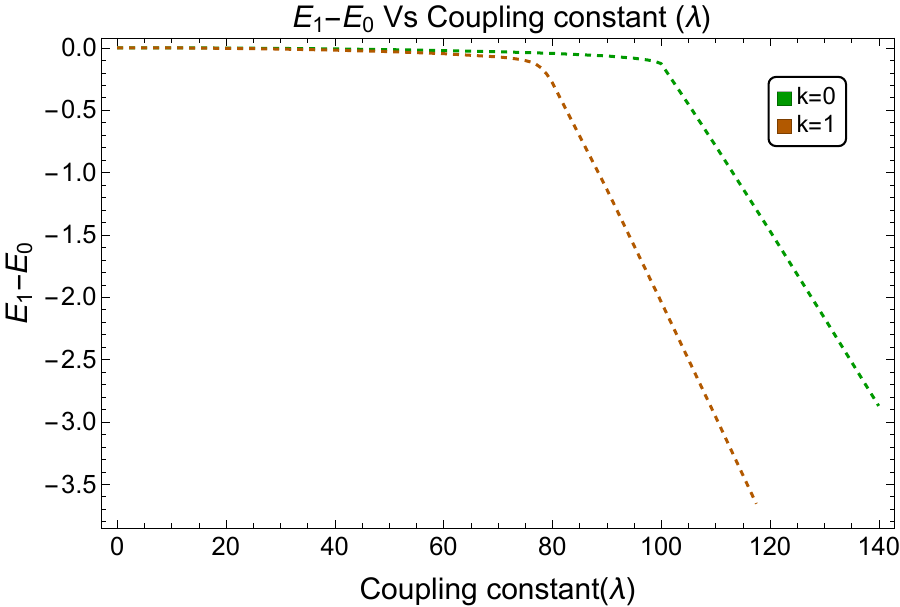}
\caption{ The negative divergence of the ground state energy with the increasing value of the coupling constant $\lambda$.}
\label{fig:the_dvg_ground_state}
\end{figure}

As seen in the Table. \ref{tab:eigenvalues_phi_4_theory_different_values_of_lambda}, the ground state energy of the system is diverging with the increasing values of $\lambda$. The negative ground state energy of the theory does not have any unphysical consequences, as this is an artifact of normal-ordering. By performing the normal-ordering, the ground state of the free scalar field theory is set to zero. Since, the actual ground state of free theory is infinite, adding an attractive interaction lowers the energy below this reference point, $0$. Consequently, the ground-state energy appearing below zero should be understood as being defined with respect to the normal-ordered vacuum, rather than as an absolute negative energy. Fig. \ref{fig:the_dvg_ground_state} depicts the negative divergence of the ground state of the $\phi^4$ theory.

\begin{figure}[hbtp]
\centering
\includegraphics[scale=0.55]{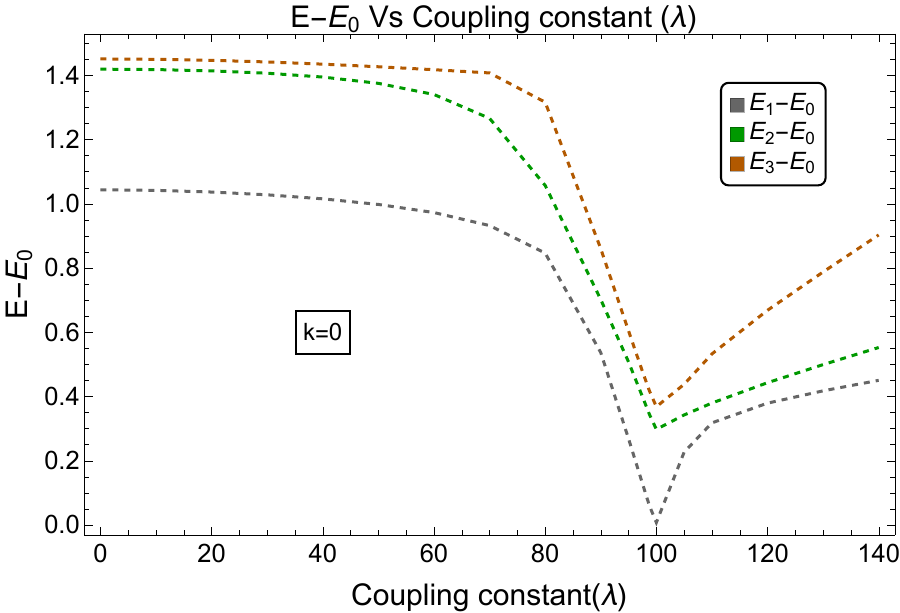}
\includegraphics[scale=0.55]{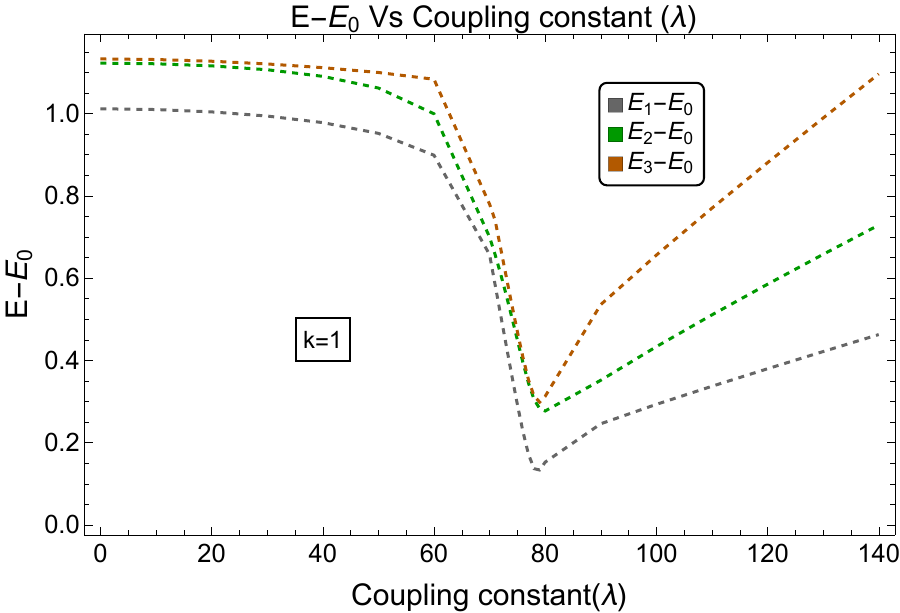}
\caption{The energy spectrum plot of the interacting $\phi^4$ theory for resolution $1$ and $2$ with the increasing values of the coupling constant $\lambda$.}
\label{fig:degenaracy_of_states}
\end{figure}

Fig. \ref{fig:degenaracy_of_states} exhibits the variation of a few excited states above the vacuum with respect to the ground state energy for resolution $k=0$ and $k=1$. For both resolutions, the ground state become almost degenerate after a particular value of $\lambda$ and then the degeneracy is lifted again. The degeneracy implies the emergence of the non-perturbative $\mathbb{Z}_2$ symmetry breaking for $m^2>0$ regime within this framework. The lifting of degeneracy is the artifact of the asymmetric nature of the basis functions. With the increment of the resolution from $0$ to $1$, the lifting is less profound because the unevenness of the distribution of the functions around the origin decreases with the increasing resolution.

\begin{figure}[H]
\centering
\includegraphics[scale=0.55]{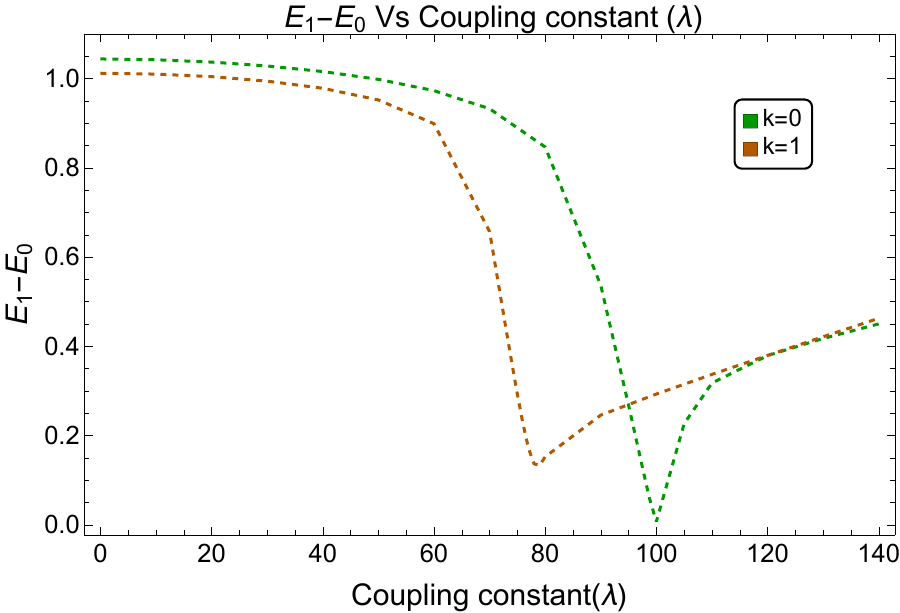}
\caption{The behaviour of the 1st excited state($E_1 - E_0$) of the $\phi^4$ theory for resolution $1$ and $2$ with the increasing value of the coupling constant $\lambda$.}
\label{fig:critical_point_for_k_1}
\end{figure}

The critical point is observed  at $\lambda_c\approx 100$ and $\lambda_c\approx 79$ for resolutions, $k=0$ and $k=1$, respectively  \ref{fig:critical_point_for_k_1}. This corresponds to the critical coupling, $g_c=\frac{\lambda_c}{24}$ to be $\sim$ 4.17, and $\sim$ 3.29, respectively. These results clearly indicate that increasing the resolution drives the critical value toward the more precise estimate $g_c \sim  2.8$
obtained using other methods \cite{PhysRevD.79.056008, Bosetti_2015, PhysRevD.99.034508, PhysRevD.88.085030, PhysRevD.106.L071501, 10.1016/j.physletb.2015.11.015, 10.1007/JHEP08(2018)148, PhysRevD.93.065014, 10.1007/JHEP05(2019)184, PhysRevD.96.065024, demiray2025systematicimprovementhamiltoniantruncation, PhysRevResearch.2.033278, PhysRevD.109.045016}. The remaining discrepancy is expected to diminish further at higher resolutions, which we plan to investigate in future work.

\section{Conclusions and outlooks}
\label{sec:conclusions_and_outlook}
In this work, we developed a Hamiltonian formulation of the $1+1$ dimensional $\phi^4$ theory using a momentum-space Daubechies wavelet basis, inspired by Wilson’s original wave-packet picture. By expanding the field operators in terms of wavelet modes characterized by resolution and translation indices, we constructed a truncated Fock space and evaluated the full Hamiltonian matrix nonperturbatively. 

The wavelet parameters, resolution index ($k$) and translation index ($n$), jointly give a clean and systematic truncation scheme as an alternate to the standard Fourier bases. Because Daubechies wavelets have compact support, the free Hamiltonian shows finite-range hopping between scaling modes (as visualised in Fig. \ref{fig:the_hopping_strength}) and for the case of the interacting $\phi^4$ theory hamiltonian, the maximum contribution comes from a smaller number of degrees of freedom, in other words the compressibility of the Hamiltonian matrix increases.

Our results demonstrate that within this formalism the critical point can be reliably tracked. Remarkably, even at very coarse resolution, the extracted critical value for the interactive $\phi^4$ theory is seen to converge toward its correct value.  This exploratory calculation gives us hope to move forward with more QFT problems. 
The optimization problem can be addressed by taking projection of the Hamiltonian to the zero momentum sector, which can reduce the computation cost, and the energy spectra for higher resolutions can also be calculated. It would be interesting to resolve this in future work. 

Looking ahead, the momentum-space wavelet framework introduced here can naturally be extended to higher dimensions. The multiresolution structure and locality properties of Daubechies wavelets make them well suited for studying more realistic interacting field theories, potentially including gauge theories and QCD-like systems. In this sense, the present work represents a first step toward a scalable, wavelet-based Hamiltonian approach to nonperturbative quantum field theory. In addition, we plan to explore Hamiltonian formulations, based on wavelets in position space, which are currently under development and will be reported in future work.

\section*{Acknowledgement}
This work was supported by the Department of Atomic Energy, Government of India, under Project Identification Number RTI 4002. The simulations were performed on the computing clusters of the Department of Theoretical Physics, TIFR. MB acknowledges support from the visitor program of the Department of Theoretical Physics, TIFR.
\appendix
\section{The $\phi^4$ overlap integral}
\label{appen:phi_4_overlap_integral}
The overlap integral appeared in Eq. (\ref{eq:phi4_hopping_term}) is,
\begin{widetext}
\begin{gather}
\label{eq:appen_phi4_hopping_term}
\Gamma^k_{q_1,q_2,q_3,q_4}=\int_{p_1+p_2+p_3+p_4=0}\frac{dp_1dp_2dp_3dp_4}{\sqrt{E_1E_2E_3E_4}}s^k_{q_1}(p_1)s^k_{q_2}(p_2)s^k_{q_3}(p_3)s^k_{q_4}(p_4),\quad\text{and}\quad E_r=p_r^2+m^2.
\end{gather}
\end{widetext}
Since the tensor is fully symmetric, it is sufficient to evaluate the integral only for the distinct index combinations with the ordering $q_1\geq q_2 \geq q_3 \geq q_4$. The results for all other cases then follow by symmetry. If each indices can take values $1,2,3,...,N$, the total number of unique combinations would be $\frac{N(N+1)(N+2)(N+3)}{24}$. Furthermore, owing to the compact support of the scaling function, the scaling functions $s^k_{q_1}$, $s^k_{q_2}$, $s^k_{q_3}$, and $s^k_{q_4}$ have nonzero value for the following momentum interval,
\begin{eqnarray}
s^k_{q_1}(p):\, q_1 \, \text{to} \,\, (q_1+\frac{5}{2^k}),\quad s^k_{q_2}(p):\, q_2 \, \text{to} \,\, (q_2+\frac{5}{2^k}),\nonumber\\
s^k_{q_3}(p):\, q_3 \, \text{to} \,\, (q_3+\frac{5}{2^k}),\quad s^k_{q_4}(p):\, q_4 \, \text{to} \,\, (q_4+\frac{5}{2^k}).\nonumber
\end{eqnarray}
Imposing momentum conservation,
\begin{eqnarray}
p_1+p_2+p_3+p_4=0,
\end{eqnarray}
leads to a constraint on the indices:
\begin{eqnarray}
-\frac{20}{2^k}\leq (q_1+q_2+q_3+q_4)\leq 0.
\end{eqnarray}
Violation of this condition implies the zero values of the integral.
\section{The interacting Hamiltonian matrix element within the Daubechies wavelet based framework}
\label{appen:the_interacting_Hamiltonian_matrix}
In this section of the appendix, we presented the construction of the interaction part of the Hamiltonian matrix element within the Daubechies scaling function based framework. The interaction part of the Hamiltonian matrix element within this framework is given by,
\begin{gather}
\mathrm{H_I}_{ss,m_1 m_2}= \frac{\lambda}{4!}\sum_{q_1,q_2,q_3,q_4}\frac{\Gamma^k_{q_1,q_2,q_3,q_4}}{(4\pi)^2}\left[\mathrm{A}^{k}_{ss,m_1m_2}-4 \times \mathrm{B}^{k}_{ss,m_1m_2}+6\times \mathrm{C}^{k}_{ss,m_1m_2}- 4 \times \mathrm{D}^{k}_{ss,m_1m_2} + \mathrm{E}^{k}_{ss,m_1m_2} \right],
\end{gather}
where,
\begin{gather}
 \mathrm{A}^{k}_{ss,m1m2}=\bra{m1^{s,k}_{-\infty}...m1^{s,k}_{\infty}}a^{s,k}_{q_1}a^{s,k}_{q_2}a^{s,k}_{q_3}a^{s,k}_{q_4}\ket{m2^{s,k}_{-\infty}...m2^{s,k}_{\infty}}\nonumber\\
=\begin{dcases*}
A^{s,k}_{q_1,q_2,q_3,q_4}\delta_{m1^{s,k}_{-\infty},m2^{s,k}_{-\infty}} \ldots \delta_{m1^{s,k}_{q_1},(m2^{s,k}_{q_1}-1)} \ldots \delta_{m1^{s,k}_{q_2},(m2^{s,k}_{q_2}-1)} \ldots \delta_{m1^{s,k}_{q_3},(m2^{s,k}_{q_3}-1)}\ldots\\
\delta_{m1^{s,k}_{q_4},(m1^{s,k}_{q_4}-1)}\ldots \delta_{m1^{s,k}_{\infty},m2^{s,k}_{\infty}},\text{ for } q_1\neq q_2 \neq q_3 \neq q_4,\\
A^{s,k}_{q_1,q_2,q_3,q_4}\delta_{m1^{s,k}_{-\infty},m2^{s,k}_{-\infty}} \ldots \delta_{m1^{s,k}_{q_1},(m2^{s,k}_{q_1}-2)}\ldots \delta_{m1^{s,k}_{q_3},(m2^{s,k}_{q_3}-1)}\ldots \delta_{m1^{s,k}_{q_4},(m2^{s,k}_{q_4}-1)}\ldots \\
\delta_{m1^{s,k}_{\infty},m2^{s,k}_{\infty}}, \text{ for } q_1= q_2 \neq q_3 \neq q_4,\\
A^{s,k}_{q_1,q_2,q_3,q_4}\delta_{m1^{s,k}_{-\infty},m2^{s,k}_{-\infty}}\ldots \delta_{m1^{s,k}_{q_1},(m2^{s,k}_{q_1}-2)}\ldots \delta_{m1^{s,k}_{q_2},(m2^{s,k}_{q_2}-1)}\ldots \delta_{m1^{s,k}_{q_4},(m2^{s,k}_{q_4}-1)}\ldots \\
\delta_{m1^{s,k}_{\infty},m2^{s,k}_{\infty}},\text{ for } q_1= q_3 \neq q_2 \neq q_4,\\
A^{s,k}_{q_1,q_2,q_3,q_4}\delta_{m1^{s,k}_{-\infty},m2^{s,k}_{-\infty}}\ldots\delta_{m1^{s,k}_{q_1},(m2^{s,k}_{q_1}-2)}\ldots \delta_{m1^{s,k}_{q_2},(m2^{s,k}_{q_2}-1)}\ldots \delta_{m1^{s,k}_{q_3},(m2^{s,k}_{q_3}-1)}\ldots\\
\delta_{m1^{s,k}_{\infty},m2^{s,k}_{\infty}},\text{ for } q_1= q_4 \neq q_2 \neq q_3,\\
A^{s,k}_{q_1,q_2,q_3,q_4}\delta_{m1^{s,k}_{-\infty},m2^{s,k}_{-\infty}}\ldots \delta_{m1^{s,k}_{q_1},(m2^{s,k}_{q_1}-1)}\ldots \delta_{m1^{s,k}_{q_2},(m2^{s,k}_{q_2}-2)}\ldots \delta_{m1^{s,k}_{q_4},(m2^{s,k}_{q_4}-1)}\ldots\\
\delta_{m1^{s,k}_{\infty},m2^{s,k}_{\infty}},\text{ for } q_2= q_3 \neq q_1 \neq q_4,\\
A^{s,k}_{q_1,q_2,q_3,q_4}\delta_{m1^{s,k}_{-\infty},m2^{s,k}_{-\infty}}\ldots \delta_{m1^{s,k}_{q_1},(m2^{s,k}_{q_1}-1)}\ldots \delta_{m1^{s,k}_{q_2},(m2^{s,k}_{q_2}-2)}\ldots \delta_{m1^{s,k}_{q_3},(m2^{s,k}_{q_3}-1)}\ldots\\
\delta_{m1^{s,k}_{\infty},m2^{s,k}_{\infty}},\text{ for } q_2= q_4 \neq q_1 \neq q_3,\\
A^{s,k}_{q_1,q_2,q_3,q_4}\delta_{m1^{s,k}_{-\infty},m2^{s,k}_{-\infty}}\ldots \delta_{m1^{s,k}_{q_1},(m2^{s,k}_{q_1}-1)}\ldots \delta_{m1^{s,k}_{q_2},(m2^{s,k}_{q_2}-1)}\ldots \delta_{m1^{s,k}_{q_3},(m2^{s,k}_{q_3}-2)}\ldots \\
\delta_{m1^{s,k}_{\infty},m2^{s,k}_{\infty}},\text{ for } q_3= q_4 \neq q_1 \neq q_2,\\
A^{s,k}_{q_1,q_2,q_3,q_4}\delta_{m1^{s,k}_{-\infty}m2^{s,k}_{-\infty}}\ldots \delta_{m1^{s,k}_{q_1},(m2^{s,k}_{q_1}-3)}\ldots \delta_{m1^{s,k}_{q_4},(m2^{s,k}_{q_4}-1)}\ldots\delta_{m1^{s,k}_{\infty},m2^{d,k}_{\infty}},\\
\text{for } q_1= q_2 = q_3 \neq q_4,\\
A^{s,k}_{q_1,q_2,q_3,q_4}\delta_{m1^{s,k}_{-\infty},m2^{s,k}_{-\infty}}\ldots \delta_{m1^{s,k}_{q_1},(m2^{s,k}_{q_1}-3)}\ldots \delta_{m1^{s,k}_{q_3},(m2^{s,k}_{q_3}-1)}\ldots \delta_{m1^{s,k}_{\infty},m2^{s,k}_{\infty}},\\
\text{for } q_1= q_2 = q_4 \neq q_3,\\
A^{s,k}_{q_1,q_2,q_3,q_4}\delta_{m1^{s,k}_{-\infty},m2^{s,k}_{-\infty}}\ldots \delta_{m1^{s,k}_{q_1},(m2^{s,k}_{q_1}-3)}\ldots \delta_{m1^{s,k}_{q_2},(m2^{s,k}_{q_2}-1)}\ldots \delta_{m1^{s,k}_{\infty},m2^{s,k}_{\infty}},\\
\text{for } q_1= q_3 = q_4 \neq q_2,\\
A^{s,k}_{q_1,q_2,q_3,q_4}\delta_{m1^{s,k}_{-\infty},m2^{s,k}_{-\infty}}\ldots \delta_{m1^{s,k}_{q_1},(m2^{s,k}_{q_1}-1)}\ldots \delta_{m1^{s,k}_{q_2},(m2^{s,k}_{q_2}-3)}\ldots \delta_{m1^{s,k}_{\infty},m2^{s,k}_{\infty}},\\
\text{for } q_2= q_3 = q_4 \neq q_1,\\
A^{s,k}_{q_1,q_2,q_3,q_4}\delta_{m1^{s,k}_{-\infty},m2^{s,k}_{-\infty}}\ldots \delta_{m1^{s,k}_{q_1},(m2^{s,k}_{q_1}-2)}\ldots \delta_{m1^{s,k}_{q_3},(m2^{s,k}_{q_3}-2)}\ldots \delta_{m1^{s,k}_{\infty},m2^{s,k}_{\infty}},\\
\text{for } q_1= q_2 \neq q_3 = q_4,\\
A^{s,k}_{q_1,q_2,q_3,q_4}\delta_{m1^{s,k}_{-\infty},m2^{s,k}_{-\infty}}\ldots \delta_{m1^{s,k}_{q_1},(m2^{s,k}_{q_1}-2)}\ldots \delta_{m1^{s,k}_{q_2},(m2^{s,k}_{q_2}-2)}\ldots \delta_{m1^{s,k}_{\infty},m2^{s,k}_{\infty}},\\
\text{for } q_1= q_3 \neq q_2 = q_4,\\
A^{s,k}_{q_1,q_2,q_3,q_4}\delta_{m1^{s,k}_{-\infty},m2^{s,k}_{-\infty}}\ldots \delta_{m1^{s,k}_{q_1},(m2^{s,k}_{q_1}-2)}\ldots \delta_{m1^{s,k}_{q_2},(m2^{s,k}_{q_2}-2)}\ldots \delta_{m1^{s,k}_{\infty},m2^{s,k}_{\infty}},\\
\text{for } q_1= q_4 \neq q_2 = q_3,\\
A^{s,k}_{q_1,q_2,q_3,q_4}=\delta_{m1^{s,k}_{-\infty},m2^{s,k}_{-\infty}}\ldots \delta_{m1^{s,k}_{q_1},(m1^{s,k}_{q_1}-4)}\ldots \delta_{m1^{s,k}_{\infty},m2^{s,k}_{\infty}},\\
\text{for }q_1= q_2 = q_3 = q_4,\\
\end{dcases*}
\end{gather}
with,
\begin{gather}
A^{s,k}_{q_1,q_2,q_3,q_4}=
\begin{dcases*}
\sqrt{m2^{s,k}_{q_1}m2^{s,k}_{q_2}m2^{s,k}_{q_3}m2^{s,k}_{q_4}} \text{ for }q_1\neq q_2 \neq q_3 \neq q_4, \\
\sqrt{(m2^{s,k}_{q_1}-1)m2^{s,k}_{q_1}m2^{s,k}_{q_3}m2^{s,k}_{q_4}}, \text{ for }q_1= q_2 \neq q_3 \neq q_4,\\
 \sqrt{(m2^{s,k}_{q_1}-1)m2^{s,k}_{q_2}m2^{s,k}_{q_1}m2^{s,k}_{q_4}}, \text{ for }q_1= q_3 \neq q_2 \neq q_4,\\
 \sqrt{(m2^{s,k}_{q_1}-1)m2^{s,k}_{q_2}m2^{s,k}_{q_3}m2^{s,k}_{q_1}}, \text{ for }q_1= q_4 \neq q_2 \neq q_3,\\
 \sqrt{m2^{s,k}_{q_1}(m2^{s,k}_{q_2}-1)m2^{s,k}_{q_2}m2^{s,k}_{q_4}}, \text{ for }q_2= q_3 \neq q_1 \neq q_4,\\
 \sqrt{m2^{s,k}_{q_1}(m2^{s,k}_{q_2}-1)m2^{s,k}_{q_3}m2^{s,k}_{q_2}}, \text{ for }q_2= q_4 \neq q_2 \neq q_3,\\ 
 \sqrt{m2^{s,k}_{q_1}m2^{s,k}_{q_2}(m2^{s,k}_{q_3}-1)m2^{s,k}_{q_3}}, \text{ for }q_3= q_4 \neq q_1 \neq q_2,\\ 
 \sqrt{(m2^{s,k}_{q_1}-2)(m2^{s,k}_{q_1}-1)m2^{s,k}_{q_1}m2^{s,k}_{q_4}}, \text{ for }q_1= q_2 = q_3 \neq q_2,\\ 
 \sqrt{(m2^{s,k}_{q_1}-2)(m2^{s,k}_{q_1}-1)m2^{s,k}_{q_3}m2^{s,k}_{q_1}}, \text{ for }q_1= q_2 = q_4 \neq q_3,\\
 \sqrt{(m2^{s,k}_{q_1}-2)m2^{s,k}_{q_2}(m2^{s,k}_{q_1}-1)m2^{s,k}_{q_1}}, \text{ for }q_1= q_3 = q_4 \neq q_2,\\
 \sqrt{m2^{s,k}_{q_1}(m2^{s,k}_{q_2}-2)(m2^{s,k}_{q_2}-1)m2^{s,k}_{q_2}}, \text{ for }q_2= q_3 = q_4 \neq q_1,\\
 \sqrt{m2^{s,k}_{q_1}(m2^{s,k}_{q_1}-1)m2^{s,k}_{q_3}(m2^{s,k}_{q_3}-1)}\text{ for }q_1= q_2 \neq q_3 = q_4,\\
 \sqrt{m2^{s,k}_{q_1}(m2^{s,k}_{q_1}-1)m2^{s,k}_{q_2}(m2^{s,k}_{q_2}-1)}\text{ for }q_1= q_3 \neq q_2 = q_4,\\
 \sqrt{m2^{s,k}_{q_1}(m2^{s,k}_{q_1}-1)m2^{s,k}_{q_2}(m2^{s,k}_{q_2}-1)}\text{ for }q_1= q_4 \neq q_2 = q_3,\\
 \sqrt{(m2^{s,k}_{q_1}-3)(m2^{s,k}_{q_1}-2)(m2^{s,k}_{q_1}-1)m2^{s,k}_{q_1}}, \text{ for }q_1= q_2= q_3 = q_4 ,
\end{dcases*}
\end{gather} 
\begin{gather}
\mathrm{B}^{k}_{ss,m_1m_2}=\bra{m1^{s,k}_{-\infty}...m1^{s,k}_{\infty}}a^{s,k\dagger}_{q_1}a^{s,k}_{q_2}a^{s,k}_{q_3}a^{s,k}_{q_4}\ket{m2^{s,k}_{-\infty}...m2^{s,k}_{\infty}}\nonumber\\
=\begin{dcases*}
B^{s,k}_{q_1,q_2,q_3,q_4}\delta_{m1^{s,k}_{-\infty},m2^{s,k}_{-\infty}} \ldots \delta_{m1^{s,k}_{q_1},(m2^{s,k}_{q_1}+1)} \ldots \delta_{m1^{s,k}_{q_2},(m2^{s,k}_{q_2}-1)} \ldots \delta_{m1^{s,k}_{q_3},(m2^{s,k}_{q_3}-1)}\ldots\\
\delta_{m1^{s,k}_{q_4},(m2^{s,k}_{q_4}-1)}\ldots \delta_{m1^{s,k}_{\infty},m2^{s,k}_{\infty}},\text{ for } q_1\neq q_2 \neq q_3 \neq q_4,\\
B^{s,k}_{q_1,q_2,q_3,q_4}\delta_{m1^{s,k}_{-\infty},m2^{s,k}_{-\infty}} \ldots \delta_{m1^{s,k}_{q_1},m2^{s,k}_{q_1}}\ldots \delta_{m1^{s,k}_{q_3},(m2^{s,k}_{q_3}-1)}\ldots \delta_{m1^{s,k}_{q_4},(m2^{s,k}_{q_4}-1)}\ldots \\
\delta_{m1^{s,k}_{\infty},m2^{s,k}_{\infty}},\text{ for } q_1= q_2 \neq q_3 \neq q_4,\\
B^{s,k}_{q_1,q_2,q_3,q_4}\delta_{m1^{s,k}_{-\infty},m2^{s,k}_{-\infty}}\ldots \delta_{m1^{s,k}_{q_1},m2^{s,k}_{q_1}}\ldots \delta_{m1^{s,k}_{q_2},(m2^{s,k}_{q_2}-1)}\ldots \delta_{m1^{s,k}_{q_4},(m2^{s,k}_{q_4}-1)}\ldots \\
\delta_{m1^{s,k}_{\infty},m2^{s,k}_{\infty}},\text{ for } q_1= q_3 \neq q_2 \neq q_4,\\
B^{s,k}_{q_1,q_2,q_3,q_4}\delta_{m1^{s,k}_{-\infty},m2^{s,k}_{-\infty}}\ldots\delta_{m1^{s,k}_{q_1},m2^{s,k}_{q_1}}\ldots \delta_{m1^{s,k}_{q_2},(m2^{s,k}_{q_2}-1)}\ldots \delta_{m1^{s,k}_{q_3},(m2^{s,k}_{q_3}-1)}\ldots\\
\delta_{m1^{s,k}_{\infty},m2^{s,k}_{\infty}},\text{ for } q_1= q_4 \neq q_2 \neq q_3,\\
B^{s,k}_{q_1,q_2,q_3,q_4}\delta_{m1^{s,k}_{-\infty},m2^{s,k}_{-\infty}}\ldots \delta_{m1^{s,k}_{q_1},(m2^{s,k}_{q_1}+1)}\ldots \delta_{m1^{s,k}_{q_2},(m2^{s,k}_{q_2}-2)}\ldots \delta_{m1^{s,k}_{q_4},(m2^{s,k}_{q_4}-1)}\ldots\\
\delta_{m1^{s,k}_{\infty},m2^{s,k}_{\infty}},\text{ for } q_2= q_3 \neq q_1 \neq q_4,\\
B^{s,k}_{q_1,q_2,q_3,q_4}\delta_{m1^{s,k}_{-\infty},m2^{s,k}_{-\infty}}\ldots \delta_{m1^{s,k}_{q_1},(m2^{s,k}_{q_1}+1)}\ldots \delta_{m1^{s,k}_{q_2},(m2^{s,k}_{q_2}-2)}\ldots \delta_{m1^{s,k}_{q_3},(m2^{s,k}_{q_3}-1)}\ldots\\
\delta_{m1^{s,k}_{\infty},m2^{s,k}_{\infty}},\text{ for }q_2= q_4 \neq q_1 \neq q_3,\\
B^{s,k}_{q_1,q_2,q_3,q_4}\delta_{m1^{s,k}_{-\infty},m2^{s,k}_{-\infty}}\ldots \delta_{m1^{s,k}_{q_1},(m2^{s,k}_{q_1}+1)}\ldots \delta_{m1^{s,k}_{q_2},(m2^{s,k}_{q_2}-1)}\ldots \delta_{m1^{s,k}_{q_3},(m2^{s,k}_{q_3}-2)}\ldots \\
\delta_{m1^{s,k}_{\infty},m2^{s,k}_{\infty}},\text{ for } q_3= q_4 \neq q_1 \neq q_2,\\
B^{s,k}_{q_1,q_2,q_3,q_4}\delta_{m1^{s,k}_{-\infty},m2^{s,k}_{-\infty}}\ldots \delta_{m1^{s,k}_{q_1},(m2^{s,k}_{q_1}-1)}\ldots \delta_{m1^{s,k}_{q_4},(m2^{s,k}_{q_4}-1)}\ldots\delta_{m1^{s,k}_{\infty},m2^{s,k}_{\infty}},\\
\text{for } q_1= q_2 = q_3 \neq q_4,\\
B^{s,k}_{q_1,q_2,q_3,q_4}\delta_{m1^{s,k}_{-\infty},m2^{s,k}_{-\infty}}\ldots \delta_{m1^{s,k}_{q_1},(m2^{s,k}_{q_1}-1)}\ldots \delta_{m1^{s,k}_{q_3},(m2^{s,k}_{q_3}-1)}\ldots \delta_{m1^{s,k}_{\infty},m2^{s,k}_{\infty}},\\
\text{for } q_1= q_2 = q_4 \neq q_3,\\
B^{s,k}_{q_1,q_2,q_3,q_4}\delta_{m1^{s,k}_{-\infty},m2^{s,k}_{-\infty}}\ldots \delta_{m1^{s,k}_{q_1},(m2^{s,k}_{q_1}-1)}\ldots \delta_{m1^{s,k}_{q_2},(m2^{s,k}_{q_2}-1)}\ldots \delta_{m1^{s,k}_{\infty},m2^{s,k}_{\infty}},\\
\text{for } q_1= q_3 = q_4 \neq q_2,\\
B^{s,k}_{q_1,q_2,q_3,q_4}\delta_{m1^{s,k}_{-\infty},m2^{s,k}_{-\infty}}\ldots \delta_{m1^{s,k}_{q_1},(m2^{s,k}_{q_1}+1)}\ldots \delta_{m1^{s,k}_{q_2},(m2^{s,k}_{q_2}-3)}\ldots \delta_{m1^{s,k}_{\infty},m2^{s,k}_{\infty}},\\
\text{for } q_2= q_3 = q_4 \neq q_1,\\
B^{s,k}_{q_1,q_2,q_3,q_4}\delta_{m1^{s,k}_{-\infty},m2^{s,k}_{-\infty}}\ldots \delta_{m1^{s,k}_{q_1},m2^{s,k}_{q_1}}\ldots \delta_{m1^{s,k}_{q_3},(m2^{s,k}_{q_3}-2)}\ldots \delta_{m1^{s,k}_{\infty},m2^{s,k}_{\infty}},\\
\text{for } q_1= q_2 \neq q_3 = q_4,\\
B^{s,k}_{q_1,q_2,q_3,q_4}\delta_{m1^{s,k}_{-\infty},m2^{s,k}_{-\infty}}\ldots \delta_{m1^{s,k}_{q_1},m2^{s,k}_{q_1}}\ldots \delta_{m1^{s,k}_{q_2},(m2^{s,k}_{q_2}-1)}\ldots \delta_{m1^{s,k}_{\infty},m2^{s,k}_{\infty}},\\
\text{for } q_1 = q_3 \neq q_2 = q_4,\\
B^{s,k}_{q_1,q_2,q_3,q_4}\delta_{m1^{s,k}_{-\infty},m2^{s,k}_{-\infty}}\ldots \delta_{m1^{s,k}_{q_1},m2^{s,k}_{q_1}}\ldots \delta_{m1^{s,k}_{q_2},(m2^{s,k}_{q_2}-2)}\ldots \delta_{m1^{s,k}_{\infty},m2^{s,k}_{\infty}},\\
\text{for } q_1= q_4 \neq q_2 = q_3,\\
B^{s,k}_{q_1,q_2,q_3,q_4}\delta_{m1^{s,k}_{-\infty},m2^{s,k}_{-\infty}}\ldots \delta_{m1^{s,k}_{q_1},(m1^{s,k}_{q_1}-2)}\ldots \delta_{m1^{s,k}_{\infty},m2^{s,k}_{\infty}},\\
\text{for }q_1= q_2 = q_3 = q_4,\\
\end{dcases*}
\end{gather}
with,
\begin{gather}
B^k_{q_1,q_2,q_3,q_4}=
\begin{dcases*}
\sqrt{(m2^{s,k}_{q_1}+1)m2^{s,k}_{q_2}m2^{s,k}_{q_3}m2^{s,k}_{q_4}}, \text{ for }q_1\neq q_2 \neq q_3 \neq q_4, \\
\sqrt{m2^{s,k}_{q_1}m2^{s,k}_{q_1}m2^{s,k}_{q_3}m2^{s,k}_{q_4}}, \text{ for }q_1= q_2 \neq q_3 \neq q_4,\\
 \sqrt{m2^{s,k}_{q_1}m2^{s,k}_{q_2}m2^{s,k}_{q_1}m2^{s,k}_{q_4}}, \text{ for }q_1= q_3 \neq q_2 \neq q_4,\\
 \sqrt{m2^{s,k}_{q_1}m2^{s,k}_{q_2}m2^{s,k}_{q_3}m2^{s,k}_{q_1}}, \text{ for }q_1= q_4 \neq q_2 \neq q_3,\\
 \sqrt{(m2^{s,k}_{q_1}+1)(m2^{s,k}_{q_2}-1)m2^{s,k}_{q_2}m2^{s,k}_{q_4}}, \text{ for }q_2= q_3 \neq q_1 \neq q_4,\\
 \sqrt{(m2^{s,k}_{q_1}+1)(m2^{s,k}_{q_2}-1)m2^{s,k}_{q_3}(m2^{s,k}_{q_2})}, \text{ for }q_2= q_4 \neq q_2 \neq q_3,\\ 
 \sqrt{(m2^{s,k}_{q_1}+1)(m2^{s,k}_{q_2})(m2^{s,k}_{q_3}-1)m2^{s,k}_{q_3}}, \text{ for }q_3= q_4 \neq q_1 \neq q_2,\\ 
 \sqrt{(m2^{s,k}_{q_1}-1)(m2^{s,k}_{q_1}-1)(m2^{s,k}_{q_1})m2^{s,k}_{q_4}}, \text{ for }q_1= q_2 = q_3 \neq q_4,\\ 
 \sqrt{(m2^{s,k}_{q_1}-1)(m2^{s,k}_{q_1}-1)m3(q_3)m2^{s,k}_{q_1}}, \text{ for }q_1= q_2 = q_4 \neq q_3,\\
 \sqrt{(m2^{s,k}_{q_1}-1)m2^{s,k}_{q_2}(m2^{s,k}_{q_1}-1)m2^{s,k}_{q_1}}, \text{ for }q_1= q_3 = q_4 \neq q_2,\\
 \sqrt{(m2^{s,k}_{q_1}+1)(m2^{s,k}_{q_2}-2)(m2^{s,k}_{q_2}-1)m2^{s,k}_{q_2}}, \text{ for }q_2= q_3 = q_4 \neq q_1,\\
 \sqrt{m2^{s,k}_{q_1}m2^{s,k}_{q_1}(m2^{s,k}_{q_3}-1)(m2^{s,k}_{q_3}-2)}, \text{ for }q_1= q_2 \neq q_3 = q_4,\\
 \sqrt{m2^{s,k}_{q_1}m2^{s,k}_{q_1}(m2^{s,k}_{q_2}-1)(m2^{s,k}_{q_2}-2)}, \text{ for }q_1= q_3 \neq q_2= q_4,\\
 \sqrt{m2^{s,k}_{q_1}m2^{s,k}_{q_1}(m2^{s,k}_{q_3}-1)(m2^{s,k}_{q_3}-2)}, \text{ for }q_1= q_4 \neq q_2 = q_3,\\
 \sqrt{(m2^{s,k}_{q_1}-2)(m2^{s,k}_{q_1}-2)(m2^{s,k}_{q_1}-1)m2^{s,k}_{q_1}}, \text{ for }q_1= q_2= q_3 = q_4,
\end{dcases*}
\end{gather}  
\begin{gather}
\mathrm{C}^{k}_{ss,m_1m_2}=\bra{m1^{s,k}_{-\infty}...m1^{s,k}_{\infty}}a^{s,k\dagger}_{q_1}a^{s,k\dagger}_{q_2}a^{s,k}_{q_3}a^{s,k}_{q_4}\ket{m2^{s,k}_{-\infty}...m2^{s,k}_{\infty}}\nonumber\\
=\begin{dcases*}
C^{s,k}_{q_1,q_2,q_3,q_4}\delta_{m1^{s,k}_{-\infty},m2^{s,k}_{-\infty}} \ldots \delta_{m1^{s,k}_{q_1},(m2^{s,k}_{q_1}+1)} \ldots \delta_{m1^{s,k}_{q_2},(m2^{s,k}_{q_2}+1)} \ldots \delta_{m1^{s,k}_{q_3},(m2^{s,k}_{q_3}-1)}\ldots\\
\delta_{m1^{s,k}_{q_4},(m1^{s,k}_{q_4}-1)}\ldots \delta_{m1^{s,k}_{\infty},m2^{s,k}_{\infty}}, \text{ for } q_1\neq q_2 \neq q_3 \neq q_4,\\
C^{s,k}_{q_1,q_2,q_3,q_4}\delta_{m1^{s,k}_{-\infty},m2^{s,k}_{-\infty}} \ldots \delta_{m1^{s,k}_{q_1},(m2^{s,k}_{q_1}+2)}\ldots \delta_{m1^{s,k}_{q_3},(m2^{s,k}_{q_3}-1)}\ldots \delta_{m1^{s,k}_{q_4},(m2^{s,k}_{q_4}-1)}\ldots \\
\delta_{m1^{s,k}_{\infty},m2^{s,k}_{\infty}}, \text{ for } q_1= q_2 \neq q_3 \neq q_4,\\
C^{s,k}_{q_1,q_2,q_3,q_4}\delta_{m1^{s,k}_{-\infty},m2^{s,k}_{-\infty}}\ldots \delta_{m1^{s,k}_{q_1},m2^{s,k}_{q_1}}\ldots \delta_{m1^{s,k}_{q_2},(m2^{s,k}_{q_2}+1)}\ldots \delta_{m1^{s,k}_{q_4},(m2^{s,k}_{q_4}-1)}\ldots \\
\delta_{m1^{s,k}_{\infty},m2^{s,k}_{\infty}}, \text{ for } q_1= q_3 \neq q_2 \neq q_4,\\
C^{s,k}_{q_1,q_2,q_3,q_4}\delta_{m1^{s,k}_{-\infty},m2^{s,k}_{-\infty}}\ldots\delta_{m1^{s,k}_{q_1},m2^{s,k}_{q_1}}\ldots \delta_{m1^{s,k}_{q_2},(m2^{s,k}_{q_2}+1)}\ldots \delta_{m1^{s,k}_{q_3},(m2^{s,k}_{q_3}-1)}\ldots\\
\delta_{m1^{s,k}_{\infty},m2^{s,k}_{\infty}},\text{ for } q_1= q_4 \neq q_2 \neq q_3,\\
C^{s,k}_{q_1,q_2,q_3,q_4}\delta_{m1^{s,k}_{-\infty},m2^{s,k}_{-\infty}}\ldots \delta_{m1^{s,k}_{q_1},(m2^{s,k}_{q_1}+1)}\ldots \delta_{m1^{s,k}_{q_2},m2^{s,k}_{q_2}}\ldots \delta_{m1^{s,k}_{q_4},(m2^{s,k}_{q_4}-1)}\ldots\\
\delta_{m1^{s,k}_{\infty},m2^{s,k}_{\infty}},\text{ for }q_2= q_3 \neq q_1 \neq q_4,\\
C^{s,k}_{q_1,q_2,q_3,q_4}\delta_{m1^{s,k}_{-\infty},m2^{s,k}_{-\infty}}\ldots \delta_{m1^{s,k}_{q_1},(m2^{s,k}_{q_1}+1)}\ldots \delta_{m1^{s,k}_{q_2},m2^{s,k}_{q_2}}\ldots \delta_{m1^{s,k}_{q_3},(m2^{s,k}_{q_3}-1)}\ldots\\
\delta_{m1^{s,k}_{\infty},m2^{s,k}_{\infty}},\text{ for }q_2= q_4 \neq q_1 \neq q_3,\\
C^{s,k}_{q_1,q_2,q_3,q_4}\delta_{m1^{s,k}_{-\infty},m2^{s,k}_{-\infty}}\ldots \delta_{m1^{s,k}_{q_1},(m2^{s,k}_{q_1}+1)}\ldots \delta_{m1^{s,k}_{q_2},(m2^{s,k}_{q_2}+1)}\ldots \delta_{m1^{s,k}_{q_3},(m2^{s,k}_{q_3}-2)}\ldots \\
\delta_{m1^{s,k}_{\infty},m2^{s,k}_{\infty}}, \text{ for }q_3= q_4 \neq q_1 \neq q_2,\\
C^{s,k}_{q_1,q_2,q_3,q_4}\delta_{m1^{s,k}_{-\infty},m2^{s,k}_{-\infty}}\ldots \delta_{m1^{s,k}_{q_1},(m2^{s,k}_{q_1}+1)}\ldots \delta_{m1^{s,k}_{q_4},(m2^{s,k}_{q_4}-1)}\ldots\delta_{m1^{s,k}_{\infty},m2^{s,k}_{\infty}},\\
\text{for }q_1= q_2 = q_3 \neq q_4,\\
C^{s,k}_{q_1,q_2,q_3,q_4}\delta_{m1^{s,k}_{-\infty},m2^{s,k}_{-\infty}}\ldots \delta_{m1^{s,k}_{q_1},(m2^{s,k}_{q_1}+1)}\ldots \delta_{m1^{s,k}_{q_3},(m2^{s,k}_{q_3}-1)}\ldots \delta_{m1^{s,k}_{\infty},m2^{s,k}_{\infty}},\\
\text{for } q_1= q_2 = q_4 \neq q_3,\\
C^{s,k}_{q_1,q_2,q_3,q_4}\delta_{m1^{s,k}_{-\infty},m2^{s,k}_{-\infty}}\ldots \delta_{m1^{s,k}_{q_1},(m2^{s,k}_{q_1}-1)}\ldots \delta_{m1^{s,k}_{q_2},(m2^{s,k}_{q_2}+1)}\ldots \delta_{m1^{s,k}_{\infty},m2^{s,k}_{\infty}},\\
\text{for } q_1= q_3 = q_4 \neq q_2,\\
C^{s,k}_{q_1,q_2,q_3,q_4}=\delta_{m1^{s,k}_{-\infty},m2^{s,k}_{-\infty}}\ldots \delta_{m1^{s,k}_{q_1},(m2^{s,k}_{q_1}+1)}\ldots \delta_{m1^{s,k}_{q_2},(m2^{s,k}_{q_2}-1)}\ldots \delta_{m1^{s,k}_{\infty},m2^{s,k}_{\infty}},\\
\text{for } q_2= q_3 = q_4 \neq q_1,\\
C^{s,k}_{q_1,q_2,q_3,q_4}=\delta_{m1^{s,k}_{-\infty},m2^{s,k}_{-\infty}}\ldots \delta_{m1^{s,k}_{q_1},(m2^{s,k}_{q_1}+2)}\ldots \delta_{m1^{s,k}_{q_3},(m2^{s,k}_{q_3}-2)}\ldots \delta_{m1^{s,k}_{\infty},m2^{s,k}_{\infty}},\\
\text{for } q_1= q_2 \neq q_3 = q_4,\\
C^{s,k}_{q_1,q_2,q_3,q_4}=\delta_{m1^{s,k}_{-\infty},m2^{s,k}_{-\infty}}\ldots \delta_{m1^{s,k}_{q_1},m2^{s,k}_{q_1}}\ldots \delta_{m1^{s,k}_{q_2},m2^{s,k}_{q_2}}\ldots \delta_{m1^{s,k}_{\infty},m2^{s,k}_{\infty}},\\
\text{for } q_1= q_3 \neq q_2 = q_4,\\
C^{s,k}_{q_1,q_2,q_3,q_4}=\delta_{m1^{s,k}_{-\infty},m2^{s,k}_{-\infty}}\ldots \delta_{m1^{s,k}_{q_1},m2^{s,k}_{q_1}}\ldots \delta_{m1^{s,k}_{q_2},m2^{s,k}_{q_2}}\ldots \delta_{m1^{s,k}_{\infty},m2^{s,k}_{\infty}},\\
\text{for } q_1= q_4 \neq q_2 = q_3,\\
C^{s,k}_{q_1,q_2,q_3,q_4}\delta_{m1^{s,k}_{-\infty},m2^{s,k}_{-\infty}}\ldots \delta_{m1^{s,k}_{q_1},m1^{s,k}_{q_1}}\ldots \delta_{m1^{s,k}_{\infty},m2^{s,k}_{\infty}},\\
\text{for }q_1= q_2 = q_3 = q_4,\\
\end{dcases*}
\end{gather}
with,
\begin{gather}
C^{s,k}_{q_1,q_2,q_3,q_4}=
\begin{dcases*}
\sqrt{(m2^{s,k}_{q_1}+1)(m2^{s,k}_{q_2}+1)m2^{s,k}_{q_3}m2^{s,k}_{q_4}}, \text{ for }q_1\neq q_2 \neq q_3 \neq q_4,\\
\sqrt{(m2^{s,k}_{q_1}+2)(m2^{s,k}_{q_1}+1)m2^{s,k}_{q_3}m2^{s,k}_{q_4}}, \text{ for }q_1= q_2 \neq q_3 \neq q_4,\\
\sqrt{m2^{s,k}_{q_1}(m2^{s,k}_{q_2}+1)m2^{s,k}_{q_1}m2^{s,k}_{q_4}}, \text{ for }q_1= q_3 \neq q_2 \neq q_4,\\
\sqrt{m2^{s,k}_{q_1}(m2^{s,k}_{q_2}+1)m2^{s,k}_{q_3}m2^{s,k}_{q_1}}, \text{ for }q_1= q_4 \neq q_2 \neq q_3,\\
\sqrt{(m2^{s,k}_{q_1}+1)m2^{s,k}_{q_2}m2^{s,k}_{q_2}m2^{s,k}_{q_4}}, \text{ for }q_2= q_3 \neq q_1 \neq q_4,\\
\sqrt{(m2^{s,k}_{q_1}+1)m2^{s,k}_{q_2}m2^{s,k}_{q_3}m2^{s,k}_{q_2}}, \text{ for }q_2= q_4 \neq q_1 \neq q_3,\\
\sqrt{(m2^{s,k}_{q_1}+1)(m2^{s,k}_{q_2}+1)(m2^{s,k}_{q_3}-1)m2^{s,k}_{q_3}}, \text{ for }q_3= q_4 \neq q_1 \neq q_2,\\
\sqrt{(m2^{s,k}_{q_1}+1)m2^{s,k}_{q_1}m2^{s,k}_{q_1}m2^{s,k}_{q_4}}, \text{ for }q_1= q_2 = q_3 \neq q_4,\\
\sqrt{(m2^{s,k}_{q_1}+1)m2^{s,k}_{q_1}m2^{s,k}_{q_3}m2^{s,k}_{q_1}}, \text{ for }q_1= q_2 = q_4 \neq q_3,\\
\sqrt{(m2^{s,k}_{q_1}-1)(m2^{s,k}_{q_2}+1)(m2^{s,k}_{q_1}-1)m2^{s,k}_{q_1}}, \text{ for }q_1= q_3 = q_4 \neq q_2,\\
\sqrt{(m2^{s,k}_{q_1}+1)(m2^{s,k}_{q_2}-1)(m2^{s,k}_{q_2}-1)m2^{s,k}_{q_2}}, \text{ for }q_2= q_3 = q_4 \neq q_1,\\
\sqrt{(m2^{s,k}_{q_1}+2)(m2^{s,k}_{q_2}+1)(m2^{s,k}_{q_3}-1)m2^{s,k}_{q_3}}, \text{ for }q_1 = q_2 \neq q_3 = q_4,\\
\sqrt{m2^{s,k}_{q_1}m2^{s,k}_{q_1}m2^{s,k}_{q_2}m2^{s,k}_{q_2}}, \text{ for }q_1 = q_3 \neq q_2 = q_4,\\
\sqrt{m2^{s,k}_{q_1}m2^{s,k}_{q_1}m2^{s,k}_{q_2}m2^{s,k}_{q_2}}, \text{ for }q_1 = q_4 \neq q_2 = q_3,\\
\sqrt{m2^{s,k}_{q_1}(m2^{s,k}_{q_1}-1)(m2^{s,k}_{q_1}-1)m2^{s,k}_{q_1}}, \text{ for }q_1= q_2 = q_3 = q_4,
\end{dcases*}
\end{gather}
\begin{gather}
\mathrm{D}^{k}_{ss,m_1m_2}=\bra{m1^{s,k}_{-\infty}...m1^{s,k}_{\infty}}a^{s,k\dagger}_{q_1}a^{s,k\dagger}_{q_2}a^{s,k\dagger}_{q_3}a^{s,k}_{q_4}\ket{m2^{s,k}_{-\infty}...m2^{s,k}_{\infty}}\nonumber\\
=\begin{dcases*}
D^{s,k}_{q_1,q_2,q_3,q_4}\delta_{m1^{s,k}_{-\infty},m2^{s,k}_{-\infty}} \ldots \delta_{m1^{s,k}_{q_1},(m2^{s,k}_{q_1}+1)} \ldots \delta_{m1^{s,k}_{q_2},(m2^{s,k}_{q_2}+1)} \ldots \delta_{m1^{s,k}_{q_3},(m2^{s,k}_{q_3}+1)}\ldots\\
\delta_{m1^{s,k}_{q_4},(m1^{s,k}_{q_4}-1)}\ldots \delta_{m1^{s,k}_{\infty},m2^{s,k}_{\infty}}, \text{ for } q_1\neq q_2 \neq q_3 \neq q_4,\\
D^{s,k}_{q_1,q_2,q_3,q_4}\delta_{m1^{s,k}_{-\infty},m2^{s,k}_{-\infty}} \ldots \delta_{m1^{s,k}_{q_1},(m2^{s,k}_{q_1}+2)}\ldots \delta_{m1^{s,k}_{q_3},(m2^{s,k}_{q_3}+1)}\ldots \delta_{m1^{s,k}_{q_4},(m2^{s,k}_{q_4}-1)}\ldots \\
\delta_{m1^{s,k}_{\infty},m2^{s,k}_{\infty}},\text{ for } q_1= q_2 \neq q_3 \neq q_4,\\
D^{s,k}_{q_1,q_2,q_3,q_4}\delta_{m1^{s,k}_{-\infty},m2^{s,k}_{-\infty}}\ldots \delta_{m1^{s,k}_{q_1},(m2^{s,k}_{q_1}+2)}\ldots \delta_{m1^{s,k}_{q_2},(m2^{s,k}_{q_2}+1)}\ldots \delta_{m1^{s,k}_{q_4},(m2^{s,k}_{q_4}-1)}\ldots \\
\delta_{m1^{s,k}_{\infty},m2^{s,k}_{\infty}},\text{ for } q_1= q_3 \neq q_2 \neq q_4,\\
D^{s,k}_{q_1,q_2,q_3,q_4}\delta_{m1^{s,k}_{-\infty},m2^{s,k}_{-\infty}}\ldots\delta_{m1^{s,k}_{q_1},m2^{s,k}_{q_1}}\ldots \delta_{m1^{s,k}_{q_2},(m2^{s,k}_{q_2}+1)}\ldots \delta_{m1^{s,k}_{q_3},(m2^{s,k}_{q_3}+1)}\ldots\\
\delta_{m1^{s,k}_{\infty},m2^{s,k}_{\infty}}, \text{ for } q_1= q_4 \neq q_2 \neq q_3,\\
D^{s,k}_{q_1,q_2,q_3,q_4}\delta_{m1^{s,k}_{-\infty},m2^{s,k}_{-\infty}}\ldots \delta_{m1^{s,k}_{q_1},(m2^{s,k}_{q_1}+1)}\ldots \delta_{m1^{s,k}_{q_2},(m2^{s,k}_{q_2}+2)}\ldots \delta_{m1^{s,k}_{q_4},(m2^{s,k}_{q_4}-1)}\ldots\\
\delta_{m1^{s,k}_{\infty},m2^{s,k}_{\infty}}, \text{ for } q_2= q_3 \neq q_1 \neq q_4,\\
D^{s,k}_{q_1,q_2,q_3,q_4}\delta_{m1^{s,k}_{-\infty},m2^{s,k}_{-\infty}}\ldots \delta_{m1^{s,k}_{q_1},(m2^{s,k}_{q_1}+1)}\ldots \delta_{m1^{s,k}_{q_2},m2^{s,k}_{q_2}}\ldots \delta_{m1^{s,k}_{q_3},(m2^{s,k}_{q_3}+1)}\ldots\\
\delta_{m1^{s,k}_{\infty},m2^{s,k}_{\infty}},\text{ for } q_2= q_4 \neq q_1 \neq q_3,\\
D^{s,k}_{q_1,q_2,q_3,q_4}\delta_{m1^{s,k}_{-\infty},m2^{s,k}_{-\infty}}\ldots \delta_{m1^{s,k}_{q_1},(m2^{s,k}_{q_1}+1)}\ldots \delta_{m1^{s,k}_{q_2},(m2^{s,k}_{q_2}+1)}\ldots \delta_{m1^{s,k}_{q_3},m2^{s,k}_{q_3}}\ldots \\
\delta_{m1^{s,k}_{\infty},m2^{s,k}_{\infty}},\text{ for } q_3= q_4 \neq q_1 \neq q_2,\\
D^{s,k}_{q_1,q_2,q_3,q_4}\delta_{m1^{s,k}_{-\infty},m2^{s,k}_{-\infty}}\ldots \delta_{m1^{s,k}_{q_1},(m2^{s,k}_{q_1}+3)}\ldots \delta_{m1^{s,k}_{q_4},(m2^{s,k}_{q_4}-1)}\ldots\delta_{m1^{s,k}_{\infty},m2^{s,k}_{\infty}},\\
\text{for } q_1= q_2 = q_3 \neq q_4,\\
D^{s,k}_{q_1,q_2,q_3,q_4}\delta_{m1^{s,k}_{-\infty},m2^{s,k}_{-\infty}}\ldots \delta_{m1^{s,k}_{q_1},(m2^{s,k}_{q_1}+1)}\ldots \delta_{m1^{s,k}_{q_3},(m2^{s,k}_{q_3}+1)}\ldots \delta_{m1^{s,k}_{\infty},m2^{s,k}_{\infty}},\\
\text{for } q_1= q_2 = q_4 \neq q_3,\\
D^{s,k}_{q_1,q_2,q_3,q_4}\delta_{m1^{s,k}_{-\infty},m2^{s,k}_{-\infty}}\ldots \delta_{m1^{s,k}_{q_1},(m2^{s,k}_{q_1}+1)}\ldots \delta_{m1^{s,k}_{q_2},(m2^{s,k}_{q_2}+1)}\ldots \delta_{m1^{s,k}_{\infty},m2^{s,k}_{\infty}},\\
\text{for } q_1= q_3 = q_4 \neq q_2,\\
D^{s,k}_{q_1,q_2,q_3,q_4}\delta_{m1^{s,k}_{-\infty},m2^{s,k}_{-\infty}}\ldots \delta_{m1^{s,k}_{q_1},(m2^{s,k}_{q_1}+1)}\ldots \delta_{m1^{s,k}_{q_2},(m2^{s,k}_{q_2}+1)}\ldots \delta_{m1^{s,k}_{\infty},m2^{s,k}_{\infty}},\\
\text{for } q_2= q_3 = q_4 \neq q_1,\\
D^{s,k}_{q_1,q_2,q_3,q_4}\delta_{m1^{s,k}_{-\infty},m2^{s,k}_{-\infty}}\ldots \delta_{m1^{s,k}_{q_1},(m2^{s,k}_{q_1}+2)}\ldots \delta_{m1^{s,k}_{q_3},m2^{s,k}_{q_3}}\ldots \delta_{m1^{s,k}_{\infty},m2^{s,k}_{\infty}},\\
\text{for } q_1 = q_2 \neq q_3 = q_4,\\
D^{s,k}_{q_1,q_2,q_3,q_4}\delta_{m1^{s,k}_{-\infty},m2^{s,k}_{-\infty}}\ldots \delta_{m1^{s,k}_{q_1},(m2^{s,k}_{q_1}+2)}\ldots \delta_{m1^{s,k}_{q_3},m2^{s,k}_{q_3}}\ldots \delta_{m1^{s,k}_{\infty},m2^{s,k}_{\infty}},\\
\text{for } q_1 = q_3 \neq q_2 = q_4,\\
D^{s,k}_{q_1,q_2,q_3,q_4}\delta_{m1^{s,k}_{-\infty},m2^{s,k}_{-\infty}}\ldots \delta_{m1^{s,k}_{q_1},m2^{s,k}_{q_1}}\ldots \delta_{m1^{s,k}_{q_3},(m2^{s,k}_{q_3}+2)}\ldots \delta_{m1^{s,k}_{\infty},m2^{s,k}_{\infty}},\\
\text{for } q_1 = q_4 \neq q_2 = q_3,\\
D^{s,k}_{q_1,q_2,q_3,q_4}\delta_{m1^{s,k}_{-\infty},m2^{s,k}_{-\infty}}\ldots \delta_{m1^{s,k}_{q_1},(m1^{s,k}_{q_1}+2)}\ldots \delta_{m1^{s,k}_{\infty},m2^{s,k}_{\infty}},\\
\text{for }q_1= q_2 = q_3 = q_4,\\
\end{dcases*}
\end{gather}
with,
\begin{gather}
D^k_{q_1,q_2,q_3,q_4}=
\begin{dcases*}
\sqrt{(m2^{s,k}_{q_1}+1)(m2^{s,k}_{q_2}+1)(m2^{s,k}_{q_3}+1)m2^{s,k}_{q_4}}, \text{ for }q_1\neq q_2 \neq q_3 \neq q_4,\\
\sqrt{(m2^{s,k}_{q_1}+2)(m2^{s,k}_{q_1}+1)(m2^{s,k}_{q_3}+1)m2^{s,k}_{q_4}}, \text{ for }q_1= q_2 \neq q_3 \neq q_4,\\
\sqrt{(m2^{s,k}_{q_1}+2)(m2^{s,k}_{q_2}+1)(m2^{s,k}_{q_1}+1)m2^{s,k}_{q_4}}, \text{ for }q_1= q_3 \neq q_2 \neq q_4,\\
\sqrt{m2^{s,k}_{q_1}(m2^{s,k}_{q_2}+1)(m2^{s,k}_{q_3}+1)m2^{s,k}_{q_1}}, \text{ for }q_1= q_4 \neq q_2 \neq q_3,\\
\sqrt{(m2^{s,k}_{q_1}+1)(m2^{s,k}_{q_2}+2)(m2^{s,k}_{q_2}+1)m2^{s,k}_{q_4}}, \text{ for }q_2= q_3 \neq q_1 \neq q_4,\\
\sqrt{(m2^{s,k}_{q_1}+1)m2^{s,k}_{q_2}(m2^{s,k}_{q_3}+1)m2^{s,k}_{q_2}}, \text{ for }q_2= q_4 \neq q_1 \neq q_3,\\
\sqrt{(m2^{s,k}_{q_1}+1)(m2^{s,k}_{q_2}+1)m2^{s,k}_{q_3}m2^{s,k}_{q_3}}, \text{ for }q_3= q_4 \neq q_1 \neq q_2,\\
\sqrt{(m2^{s,k}_{q_1}+3)(m2^{s,k}_{q_1}+2)(m2^{s,k}_{q_1}+1)m2^{s,k}_{q_4}}, \text{ for }q_1= q_2 = q_3 \neq q_4,\\
\sqrt{(m2^{s,k}_{q_1}+1)m2^{s,k}_{q_1}(m2^{s,k}_{q_3}+1)m2^{s,k}_{q_1}}, \text{ for }q_1= q_2 = q_4 \neq q_3,\\
\sqrt{(m2^{s,k}_{q_1}+1)(m2^{s,k}_{q_2}+1)m2^{s,k}_{q_1}m2^{s,k}_{q_1}}, \text{ for }q_1= q_3 = q_4 \neq q_2,\\
\sqrt{(m2^{s,k}_{q_1}+1)(m2^{s,k}_{q_2}+1)m2^{s,k}_{q_2}m2^{s,k}_{q_2}}, \text{ for }q_2= q_3 = q_4 \neq q_1,\\
\sqrt{(m2^{s,k}_{q_1}+2)(m2^{s,k}_{q_1}+1)m2^{s,k}_{q_3}m2^{s,k}_{q_3}}, \text{ for }q_1= q_2 \neq q_3 = q_4,\\
\sqrt{(m2^{s,k}_{q_1}+2)(m2^{s,k}_{q_1}+1)m2^{s,k}_{q_2}m2^{s,k}_{q_2}}, \text{ for }q_1= q_3 \neq q_2 = q_4,\\
\sqrt{m2^{s,k}_{q_1}m2^{s,k}_{q_1}(m2^{s,k}_{q_2}+2)(m2^{s,k}_{q_3}+1)}, \text{ for }q_1= q_4 \neq q_2 = q_3,\\
\sqrt{(m2^{s,k}_{q_1}+2)(m2^{s,k}_{q_1}+1)m2^{s,k}_{q_1}m2^{s,k}_{q_1}}, \text{ for }q_1= q_2 = q_3 = q_4,
\end{dcases*}
\end{gather}
\begin{gather}
\mathrm{E}^{k}_{ss,m_1m_2}=\bra{m1^{s,k}_{-\infty}...m1^{s,k}_{\infty}}a^{s,k\dagger}_{q_1}a^{s,k\dagger}_{q_2}a^{s,k\dagger}_{q_3}a^{s,k\dagger}_{q_4}\ket{m2^{s,k}_{-\infty}...m2^{s,k}_{\infty}}\nonumber\\
=\begin{dcases*}
E^{s,k}_{q_1,q_2,q_3,q_4}\delta_{m1^{s,k}_{-\infty},m2^{s,k}_{-\infty}} \ldots \delta_{m1^{s,k}_{q_1},(m2^{s,k}_{q_1}+1)} \ldots \delta_{m1^{s,k}_{q_2},(m2^{s,k}_{q_2}+1)} \ldots \delta_{m1^{s,k}_{q_3},(m2^{s,k}_{q_3}+1)}\ldots\\
\delta_{m1^{s,k}_{q_4},(m1^{s,k}_{q_4}+1)}\ldots \delta_{m1^{s,k}_{\infty},m2^{s,k}_{\infty}}, \text{ for } q_1\neq q_2 \neq q_3 \neq q_4,\\
E^{s,k}_{q_1,q_2,q_3,q_4}\delta_{m1^{s,k}_{-\infty},m2^{s,k}_{-\infty}} \ldots \delta_{m1^{s,k}_{q_1},(m2^{s,k}_{q_1}+2)}\ldots \delta_{m1^{s,k}_{q_3},(m2^{s,k}_{q_3}+1)}\ldots \delta_{m1^{s,k}_{q_4},(m2^{s,k}_{q_4}+1)}\ldots \\
\delta_{m1^{s,k}_{\infty},m2^{s,k}_{\infty}},\text{ for } q_1= q_2 \neq q_3 \neq q_4,\\
E^{s,k}_{q_1,q_2,q_3,q_4}\delta_{m1^{s,k}_{-\infty},m2^{s,k}_{-\infty}}\ldots \delta_{m1^{s,k}_{q_1},(m2^{s,k}_{q_1}+2)}\ldots \delta_{m1^{s,k}_{q_2},(m2^{s,k}_{q_2}+1)}\ldots \delta_{m1^{s,k}_{q_4},(m2^{s,k}_{q_4}+1)}\ldots \\
\delta_{m1^{s,k}_{\infty},m2^{s,k}_{\infty}},\text{ for } q_1= q_3 \neq q_2 \neq q_4,\\
E^{s,k}_{q_1,q_2,q_3,q_4}\delta_{m1^{s,k}_{-\infty},m2^{s,k}_{-\infty}}\ldots\delta_{m1^{s,k}_{q_1},(m2^{s,k}_{q_1}+2)}\ldots \delta_{m1^{s,k}_{q_2},(m2^{s,k}_{q_2}+1)}\ldots \delta_{m1^{s,k}_{q_3},(m2^{s,k}_{q_3}+1)}\ldots\\
\delta_{m1^{s,k}_{\infty},m2^{s,k}_{\infty}}, \text{ for } q_1= q_4 \neq q_2 \neq q_3,\\
E^{s,k}_{q_1,q_2,q_3,q_4}\delta_{m1^{s,k}_{-\infty},m2^{s,k}_{-\infty}}\ldots \delta_{m1^{s,k}_{q_1},(m2^{s,k}_{q_1}+1)}\ldots \delta_{m1^{s,k}_{q_2},(m2^{s,k}_{q_2}+2)}\ldots \delta_{m1^{s,k}_{q_4},(m2^{s,k}_{q_4}+1)}\ldots\\
\delta_{m1^{s,k}_{\infty},m2^{s,k}_{\infty}}, \text{ for } q_2= q_3 \neq q_1 \neq q_4,\\
E^{s,k}_{q_1,q_2,q_3,q_4}\delta_{m1^{s,k}_{-\infty},m2^{s,k}_{-\infty}}\ldots \delta_{m1^{s,k}_{q_1},(m2^{s,k}_{q_1}+1)}\ldots \delta_{m1^{s,k}_{q_2},(m2^{s,k}_{q_2}+2)}\ldots \delta_{m1^{s,k}_{q_3},(m2^{s,k}_{q_3}+1)}\ldots\\
\delta_{m1^{s,k}_{\infty},m2^{s,k}_{\infty}},\text{ for } q_2= q_4 \neq q_1 \neq q_3,\\
E^{s,k}_{q_1,q_2,q_3,q_4}\delta_{m1^{s,k}_{-\infty},m2^{s,k}_{-\infty}}\ldots \delta_{m1^{s,k}_{q_1},(m2^{s,k}_{q_1}+1)}\ldots \delta_{m1^{s,k}_{q_2},(m2^{s,k}_{q_2}+1)}\ldots \delta_{m1^{s,k}_{q_3},(m2^{s,k}_{q_3}+2)}\ldots \\
\delta_{m1^{s,k}_{\infty},m2^{s,k}_{\infty}},\text{ for } q_3= q_4 \neq q_1 \neq q_2,\\
E^{s,k}_{q_1,q_2,q_3,q_4}\delta_{m1^{s,k}_{-\infty},m2^{s,k}_{-\infty}}\ldots \delta_{m1^{s,k}_{q_1},(m2^{s,k}_{q_1}+3)}\ldots \delta_{m1^{s,k}_{q_4},(m2^{s,k}_{q_4}+1)}\ldots\delta_{m1^{s,k}_{\infty},m2^{s,k}_{\infty}},\\
\text{for } q_1= q_2 = q_3 \neq q_4,\\
E^{s,k}_{q_1,q_2,q_3,q_4}\delta_{m1^{s,k}_{-\infty},m2^{s,k}_{-\infty}}\ldots \delta_{m1^{s,k}_{q_1},(m2^{s,k}_{q_1}+3)}\ldots \delta_{m1^{s,k}_{q_3},(m2^{s,k}_{q_3}+1)}\ldots \delta_{m1^{s,k}_{\infty},m2^{s,k}_{\infty}},\\
\text{for } q_1= q_2 = q_4 \neq q_3,\\
E^{s,k}_{q_1,q_2,q_3,q_4}\delta_{m1^{s,k}_{-\infty},m2^{s,k}_{-\infty}}\ldots \delta_{m1^{s,k}_{q_1},(m2^{s,k}_{q_1}+3)}\ldots \delta_{m1^{s,k}_{q_2},(m2^{s,k}_{q_2}+1)}\ldots \delta_{m1^{s,k}_{\infty},m2^{s,k}_{\infty}},\\
\text{for } q_1= q_3 = q_4 \neq q_2,\\
E^{s,k}_{q_1,q_2,q_3,q_4}\delta_{m1^{s,k}_{-\infty},m2^{s,k}_{-\infty}}\ldots \delta_{m1^{s,k}_{q_1},(m2^{s,k}_{q_1}+1)}\ldots \delta_{m1^{s,k}_{q_2},(m2^{s,k}_{q_2}+3)}\ldots \delta_{m1^{s,k}_{\infty},m2^{s,k}_{\infty}},\\
\text{for } q_2= q_3 = q_4 \neq q_1,\\
E^{s,k}_{q_1,q_2,q_3,q_4}\delta_{m1^{s,k}_{-\infty},m2^{s,k}_{-\infty}}\ldots \delta_{m1^{s,k}_{q_1},(m2^{s,k}_{q_1}+2)}\ldots \delta_{m1^{s,k}_{q_3},(m2^{s,k}_{q_3}+2)}\ldots \delta_{m1^{s,k}_{\infty},m2^{s,k}_{\infty}},\\
\text{for } q_1= q_2 \neq q_3 = q_4,\\
E^{s,k}_{q_1,q_2,q_3,q_4}\delta_{m1^{s,k}_{-\infty},m2^{s,k}_{-\infty}}\ldots \delta_{m1^{s,k}_{q_1},(m2^{s,k}_{q_1}+2)}\ldots \delta_{m1^{s,k}_{q_2},(m2^{s,k}_{q_2}+2)}\ldots \delta_{m1^{s,k}_{\infty},m2^{s,k}_{\infty}},\\
\text{for } q_1= q_3 \neq q_2 = q_4,\\
E^{s,k}_{q_1,q_2,q_3,q_4}\delta_{m1^{s,k}_{-\infty},m2^{s,k}_{-\infty}}\ldots \delta_{m1^{s,k}_{q_1},(m2^{s,k}_{q_1}+2)}\ldots \delta_{m1^{s,k}_{q_2},(m2^{s,k}_{q_2}+2)}\ldots \delta_{m1^{s,k}_{\infty},m2^{s,k}_{\infty}},\\
\text{for } q_1= q_4 \neq q_2 = q_3,\\
E^{s,k}_{q_1,q_2,q_3,q_4}\delta_{m1^{s,k}_{-\infty},m2^{s,k}_{-\infty}}\ldots \delta_{m1^{s,k}_{q_1},(m1^{s,k}_{q_1}+4)}\ldots \delta_{m1^{s,k}_{\infty},m2^{s,k}_{\infty}},\\
\text{for }q_1= q_2 = q_3 = q_4,\\
\end{dcases*}
\end{gather}
with,
\begin{gather}
E^{s,k}_{q_1,q_2,q_3,q_4}=
\begin{dcases*}
\sqrt{(m2^{s,k}_{q_1}+1)(m2^{s,k}_{q_2}+1)(m2^{s,k}_{q_3}+1)(m2^{s,k}_{q_4}+1)}, \text{ for }q_1\neq q_2 \neq q_3 \neq q_4,\\
\sqrt{(m2^{s,k}_{q_1}+2)(m2^{s,k}_{q_1}+1)(m2^{s,k}_{q_3}+1)(m2^{s,k}_{q_4}+1)}, \text{ for }q_1= q_2 \neq q_3 \neq q_4,\\
\sqrt{(m2^{s,k}_{q_1}+2)(m2^{s,k}_{q_2}+1)(m2^{s,k}_{q_1}+1)(m2^{s,k}_{q_4}+1)}, \text{ for }q_1= q_3 \neq q_2 \neq q_4,\\
\sqrt{(m2^{s,k}_{q_1}+2)(m2^{s,k}_{q_2}+1)(m2^{s,k}_{q_3}+1)(m2^{s,k}_{q_1}+1)}, \text{ for }q_1= q_4 \neq q_2 \neq q_3,\\
\sqrt{(m2^{s,k}_{q_1}+1)(m2^{s,k}_{q_2}+2)(m2^{s,k}_{q_2}+1)(m2^{s,k}_{q_4}+1)}, \text{ for }q_2= q_3 \neq q_1 \neq q_4,\\
\sqrt{(m2^{s,k}_{q_1}+1)(m2^{s,k}_{q_2}+2)(m2^{s,k}_{q_3}+1)(m2^{s,k}_{q_2}+1)}, \text{ for }q_2= q_4 \neq q_1 \neq q_3,\\
\sqrt{(m2^{s,k}_{q_1}+1)(m2^{s,k}_{q_2}+1)(m2^{s,k}_{q_3}+2)(m2^{s,k}_{q_3}+1)}, \text{ for }q_3= q_4 \neq q_1 \neq q_2,\\
\sqrt{(m2^{s,k}_{q_1}+3)(m2^{s,k}_{q_1}+2)(m2^{s,k}_{q_1}+1)(m2^{s,k}_{q_4}+1)}, \text{ for }q_1= q_2 = q_3 \neq q_4,\\
\sqrt{(m2^{s,k}_{q_1}+3)(m2^{s,k}_{q_1}+2)(m2^{s,k}_{q_3}+1)(m2^{s,k}_{q_1}+1)}, \text{ for }q_1= q_2 = q_4 \neq q_3,\\
\sqrt{(m2^{s,k}_{q_1}+3)(m2^{s,k}_{q_2}+1)(m2^{s,k}_{q_1}+2)(m2^{s,k}_{q_1}+1)}, \text{ for }q_1= q_3 = q_4 \neq q_2,\\
\sqrt{(m2^{s,k}_{q_1}+1)(m2^{s,k}_{q_2}+3)(m2^{s,k}_{q_2}+2)(m2^{s,k}_{q_2}+1)}, \text{ for }q_2= q_3 = q_4 \neq q_1,\\
\sqrt{(m2^{s,k}_{q_1}+2)(m2^{s,k}_{q_1}+1)(m2^{s,k}_{q_3}+2)(m2^{s,k}_{q_3}+1)}, \text{ for }q_1= q_2 \neq q_3 = q_4,\\
\sqrt{(m2^{s,k}_{q_1}+2)(m2^{s,k}_{q_1}+1)(m2^{s,k}_{q_2}+2)(m2^{s,k}_{q_2}+1)}, \text{ for }q_1= q_3 \neq q_2 = q_4,\\
\sqrt{(m2^{s,k}_{q_1}+2)(m2^{s,k}_{q_1}+1)(m2^{s,k}_{q_2}+2)(m2^{s,k}_{q_2}+1)}, \text{ for }q_1= q_4 \neq q_2 = q_4,\\
\sqrt{(m2^{s,k}_{q_1}+4)(m2^{s,k}_{q_1}+3)(m2^{s,k}_{q_1}+2)(m2^{s,k}_{q_1}+1)}, \text{ for }q_1= q_2 = q_3 = q_4
\end{dcases*}
\end{gather}
\bibliographystyle{unsrt}
\bibliography{references}
\end{document}